\documentclass[a4paper,11pt]{article}
\usepackage{amsfonts}

\usepackage{graphicx}
\graphicspath{{Figures/}}
\usepackage{amsmath}
\usepackage{amssymb}
\usepackage{latexsym}
\usepackage[colorlinks]{hyperref}
\usepackage{color}
\usepackage{float}
\usepackage{cite}
\usepackage{makeidx}
\usepackage{colortbl}
\usepackage{csquotes}
\usepackage[squaren]{SIunits}

\usepackage[textfont={footnotesize,sf},labelfont={color=blue,bf,sf},labelsep=endash]{caption}
\usepackage[position=top,labelfont={color=blue,bf,sf}]{subfig}
\usepackage{bm}

\newcommand{\bra}{\begin{array}}
\newcommand{\era}{\end{array}}
\newcommand{\beq}{\begin{equation}}
\newcommand{\eeq}{\end{equation}}
\newcommand{\bqr}{\begin{eqnarray}}
\newcommand{\eqr}{\end{eqnarray}}

\def\BC{\bb C}
\def\_\BC{\bbi C}

\def\Tr {{\rm Tr}}

\def\( {\left(}
\def\) {\right)}
\def\no2 {{\textstyle{n\over 2}}}

\def\Tr {{\rm Tr}}

\begin{document}
\begin{titlepage}
\setcounter{page}{1}
\renewcommand{\thefootnote}{\fnsymbol{footnote}}
\begin{flushright}
\end{flushright}
\vspace{5mm}
\begin{center}
{\Large \bf { Dynamics and Redistribution  of Entanglement and Coherence in Three Time-Dependent  Coupled Harmonic Oscillators 
}}\\

\vspace{5mm}

{
{ \bf {Radouan Hab-arrih}$^{a}$, \bf Ahmed Jellal\footnote{\sf a.jellal@ucd.ac.ma}}$^{a,b}$ and
 {\bf Abdeldjalil Merdaci}$^{c}$}

\vspace{5mm}
{$^{a}$\em Laboratory of Theoretical Physics,  
Faculty of Sciences, Choua\"ib Doukkali University},\\
{\em PO Box 20, 24000 El Jadida, Morocco}

{$^{b}$\em Canadian Quantum  Research Center,
	204-3002 32 Ave Vernon, \\ BC V1T 2L7,  Canada}

{$^c$\em Physics Department, College of Science, King Faisal University,
\\
PO Box
380, Alahsa 31982, Saudi Arabia}
\vspace{30mm}

\begin{abstract}

We study the dynamics and redistribution of entanglement and coherence in three time-dependent coupled harmonic oscillators.
We resolve the Schr\"{o}dinger equation by using  time-dependent Euler rotation together with a linear quench model to obtain the state of  vacuum solution. Such state can be  translated to  the phase space picture to determine the Wigner distribution. We show that its Gaussian matrix $\mathbb{G}(t)$ can be used to directly cast the covariance matrix $\sigma(t)$. To quantify the mixedness  and entanglement of the state  one uses respectively linear    and  von Neumann entropies for three cases: fully symmetric, bi-symmetric and fully non symmetric. Then we determine the coherence, tripartite entanglement and local uncertainties and derive their dynamics. We show that the dynamics of all quantum information quantities are driven by the Ermakov modes. Finally, we use an homodyne detection to redistribute both resources of entanglement and coherence. 

\vspace{30mm}

\end{abstract}
\end{center}
\end{titlepage}

\section{Introduction}

Quantum information has reached an important milestones in the last decade \cite{R13}. Entanglement and coherence  are   the most amazing quantum world resources, which allow quantum technologies to go beyond the classical scenarios. In particular, pioneering protocols like quantum cryptography \cite{R15} and quantum teleportation \cite{R16} have been demonstrated in several experiments with a variety of quantum hardware, and entered a novel of commercialization \cite{R17}.  Traditionally, the quantum information protocols are mainly based on two approaches. The first one is digital, in which 
the information is encoded in discrete systems with finite numbers of degrees freedom likes qubit and qutrit. As a physical realizations one can use for example  polarisation of photons, nuclear spins in molecules. The second one is  analog, in which  the  information is encoded in infinite number of degrees of freedom called continuous variables \cite{R18}.  For example light quadratures, collective magnetic moments and harmonic oscillators are typical implementations.

Time-dependent harmonic oscillators (TDHO) have attracted  remarkable interest in different scientific branches thanks to their power to describe the dynamics of many physical systems in the vicinity of equilibrium 
\cite{R19,R20}. {Note that the time-dependence in  TDHO is carried by  their  frequencies and coupling parameters,
	which leads for instance to an optimal control of entanglement even at high temperature and for a
	large simulation time \cite{Roversi}}. TDHO has a prominent role in  trapping different objects like atoms \cite{R43} and molecules   \cite{R44, R41}, biological systems \cite{Molski2010}, viruses and bacteria \cite{R45,R46}. TDHO  is widely used in shortcuts to adiabacity  \cite{R42,R36} and 
to investigate  the dynamics  impact on entanglement and other related quantum quantities \cite{R5, R3,R4, R21} as well as to describe the quantum dynamics of a charged particle  in a time-varying magnetic field  \cite{R2}. For 3D model, Merdaci and Jellal  \cite{R31} studied three coupled time-independent anisotropic oscillators such that the associated Hamiltonian was diagonalized  using the $SU(3)$ unitary transformation. This allowed them to give the amount of entanglement and purity encoded in the corresponding ground state. Based on 
 sudden quenched model (SQM)
 the 3D time-dependent version  with specific relation between coupling  parameters and frequencies  was treated  \cite{R4}.
 The frequencies and coupling are abruptly changed
 in SQM, which  makes the resolution of time-dependent Schr\"{o}dinger equation (TDSE)   more economic with respect to unitary transformations used to decouple the Hamiltonian.
 However,  this model does not take into account the dynamical effect of the rotation used to decouple the Hamiltonian.
 
The Gaussian states are prototypical states in quantum optics and quantum information processing arena,  which is due to the fact that those objects have a wonderful mathematical background. Such states are completely described by the first and second moments that is the covariance matrix (CM). The first moment is not important in entanglement theory because it can be removed by local operators \cite{R14}, but it matters in coherence theory \cite{R25}. The Wigner formalism   has an paramount role in quantum information theory, which is due to the smooth behaviour of Wigner distribution under unitary transformations \cite{R8,R7}.   The partial tracing procedure  removes head scratching with complicated integrals \cite{R18} and provides the possibility to nice geometrical interpretations of some  important  mathematical criteria of separability. For example the  positive partial transpose (PPT)  criteria for CV in bipartite states act as a mirror reflections on Wigner distribution 
\cite{R7}.

The Heisenberg uncertainty plays a paramount role in quantum description  and  is at the core of quantum physics. It
presents a key of the discrepancies between classical and quantum systems \cite{R37}, then its violation implies the classicality of the states. 
In fact, if the state is not 
entangled then it is  not necessarily classical \cite{R18}, which is obvious because the quantum correlations always exceed entanglement amount except the case of pure states that  are equal. 
The coherence resource has a deeper meaning on the nature of the quantum world. This comes because it is directly linked to the superposition principle, which is the generator of the amazing quantum world (quantum interference, entanglement, $\cdots$)
\cite{R38}.

We will study the dynamics of entanglement, purities, uncertainties, correlations and coherence encoded on the ground state of  three coupled time-dependent harmonic oscillator  in the framework of SQM.
To achieve this goal, we resolve the TDSE in  general case by assuming that the parameters are arbitrary independent, in contrast with \cite{R4, R30}. {Such consideration allows for example to investigate the effect of the virtual photons in the generation of entanglement beyond the resonance regime \cite{ref1} and also to inspect the effect of quantum synchronization for robustness of quantum correlations \cite{ref2}}.
By using  the time dependent Euler rotation together with linear SQM, we diagonalize the Hamiltonian of our system.  The quantum information  will be derived  in the phase space picture \cite{R18}, because  the ground state is Gaussian.
Then, 
 we derive the associate Wigner distribution  and subsequently compute CM $\sigma(t)$,  which encodes all quantum information including coherence because the state is centered (without first moment).
 The global ground state is pure $\det\sigma(t)=1$, then we can hire the von Neumann entropy $S_{V}$ as  legitimate quantifier of three  bipartite entanglement \cite{R18}.  The knowledge  of marginal purities $P_{m}$ $(m=A,B,C)$ of three modes suffice to quantify entanglement  of bipartitions $(i|jk)$  where $i,j,k$ are different elements in $\left\lbrace A,B,C\right\rbrace $ \cite{R11}. 
 The dynamics of entanglement is a very important way to produce entanglement, we initially prepare the state $\Psi(t=0)$ then  choose the driven coupling and frequencies trajectories $\tau:\left(C_{ij}(0),\omega_{0}\right) \rightarrow \left(C_{ij}(f),\omega_{f}\right)$ to control the dynamics in order to generate an optimal  resource of entanglement.
 
The present paper is organized as follows. In section \ref{sect1}, we diagonalize the Hamiltonian by using a time dependent  Euler rotation together with a specific linear choice of coupling and frequencies. { 
In section \ref{Sec3},  we establish the linear quenched model for our system and obtain the  solutions of energy spectrum}. In section \ref{sect2},  we compute the covariance from the Wigner distribution. In section \ref{sect3} we compute the entanglement and mixedness in three cases symmetric, bi-symmetric and fully non symmetric states. In section \ref{sect4}, we derive the dynamics of uncertainties, global and local coherences in the case of symmetric state. In section \ref{sect5}, we use an homodyne detection to redistribute the resources of Gaussian state. Finally, we conclude our results.

\section{Hamiltonian formalism and transformation \label{sect1}}

{To achieve our goal, let us consider  three coupled Harmonic oscillators with  unit masses 
	described by  the following Hamiltonian 
\begin{equation}
H=\frac{1}{2}\left( \hat{p}_{1}^{2}+\hat{p}_{2}^{2}+\hat{p}_{3}^{2}\right)+\frac{1}{2}\sum_{i=1}^{3}\omega_{i}^{2}(t)\hat{x}_{i}^{2}-C_{13}(t)\hat{x}_{1}\hat{x}_{3}-C_{12}(t)\hat{x}_{1}\hat{x}_{2}-C_{23}(t)\hat{x}_{2}\hat{x}_{3}\label{er}
\end{equation}
such that the involved angular frequencies $\omega_{i}(t)$ and coupling parameters $ C_{ij}(t) $ are taken to be 
arbitrarly time-dependent, with $ i,j=1,2,3 $. 
Note that, the assumption of unit masses can be achieved under a simple unitary transformations, more details can be found in \cite{R33,Moya}.

We can diagonalize (\ref{er}) by introducing a convenient unitary transformation whose  operator is given in terms of 
$(\psi(t),\theta(t),\phi(t))$ are the Euler angles 
and   the angular moment operators
$\hat{L}_{k}=(\hat{x}_{i}\hat{p}_{j}-\hat{x}_{j}\hat{p}_{i})$  fulfill the algebra $\left[\hat{L}_{i},\hat{L}_{j}\right]=i\epsilon_{ijk}\hat{L}_{k}$
 \cite{R6}, such as 
\begin{equation}
\mathbb{U}\equiv\mathbb{U}(\psi(t),\theta(t),\phi(t))=e^{-i\psi(t)\hat{L}_{3}}e^{-i\theta(t)\hat{L}_{2}}e^{-i\phi(t)\hat{L}_{3}}.
\end{equation}
Consequently, the Hamiltonian \eqref{er}} and  associated wave function $\Psi(x_{1},x_{2},x_{3}:t)$   transform as
\begin{eqnarray}
&&H^{'}=\mathbb{U}(t)H\mathbb{U}^{-1}(t)-i\mathbb{U}(t)\frac{\partial \mathbb{U}^{-1}(t)}{\partial t}\label{e3}\\
&&
\Psi(x_{1},x_{2},x_{3}:t)=\mathbb{U}^{-1}\Psi^{'}(x_{1},x_{2},x_{3}:t)
\end{eqnarray}
where the second term in $H'$ is 
\begin{equation}
{i\mathbb{U}(t)\frac{\partial \mathbb{U}^{-1}(t)}{\partial t}=-\sum\limits_{i=1}^{3}a_{i}(t)\cdot \hat{L}_{i}\label{E5}}
\end{equation} 
and 
time-dependent Euler frequencies are given by
\begin{eqnarray}
 a_{1}(t)&=&-\dot{\theta}\cos(\psi)+\dot{\phi}\cos(\psi)\sin(\theta)\\
a_{2}(t)&=&\dot{\theta}\cos(\phi)-\dot{\phi}\sin(\psi)\sin(\theta)\\
a_{3}(t)&=&\dot{\psi}+\dot{\phi}\cos(\theta). 
\end{eqnarray}
Note that the transformation rotates the spacial coordinates $\vec x$
as  
$\mathbb{U}\vec x \mathbb{U}^{-1}=\mathbb{R}\vec x$,
where $\mathbb{R}$ is the real $3 \times 3$ orthogonal time-dependent
matrix with unit determinant corresponding to $\mathbb{U}$.
To obtain  $H^{'}$, we replace the angular momentum operators by three dimensional matrix representation, namely  $\left(L_{i}\right)_{jk}=-i\epsilon_{ijk}$ where $\epsilon_{ijk}$ is the Levi-Civita antisymmetric tensor. Then the time dependent Eulerian matrix $\mathbb{R}(\phi,\theta,\psi)=\mathbb{R}_{1}(\psi)\mathbb{R}_{2}(\theta)\mathbb{R}_{3}(\phi)$ is given by 
\begin{eqnarray}
\mathbb{R}(\phi,\theta,\psi)
=
\begin{pmatrix}
c_{\psi}c_{\theta}c_{\phi}-s_{\psi}s_{\phi} & c_{\psi}c_{\theta}s_{\phi}+s_{\psi}c_{\phi} & c_{\psi}s_{\theta} \\ 
-s_{\psi}c_{\theta}c_{\phi}-c_{\psi}s_{\phi} & -s_{\psi}c_{\theta}s_{\phi}+c_{\psi}c_{\phi} & -s_{\psi}s_{\theta} \\ 
-s_{\theta}c_{\phi} & -s_{\theta}s_{\phi} & c_{\theta}
\end{pmatrix}  =\left(\mathbb{R}_{ij}(t)\right)_{1\leqslant i,j \leqslant 3}\label{E}
\end{eqnarray}
where we have set $(s_{\eta},c_{\eta})\equiv (\sin\eta,\cos\eta)$,  $\eta\in\lbrace\theta,\phi,\psi\rbrace$
and  
the three generators $R_{i}(\eta)$ ($i=1,2,3$) are elements of the group $SO(3,\mathbb{R})$. 
We can easily obtain the relation
\begin{equation}
 {\left(\mathbb{R}^{T}_{\psi,\theta,\phi}\dot{\mathbb{R}}_{\psi,\theta,\phi}\right)_{ij}=
 \epsilon_{ijk}a_{k}}
\end{equation}
Combining all to write
 (\ref{e3})  as 
\begin{equation}
H^{'}(y_{1},y_{2},y_{3},t)=\dfrac{1}{2}\vec{\pi}^{T}\vec{\pi}+\frac{1}{2}\vec{y}^{T}\mathbb{R}^{T}\mathbb{C}(t)\mathbb{R}\vec{y}+\sum_{i=1}^{3}a_{i}(t)\hat{L}_{i} \label{8}
\end{equation}where the angular momenta operators now are  $ \hat{L}_{k}= y_{i}\pi_{j}-y_{j}\pi_{i}$, new variables 
 $\vec{y}=(y_{1},y_{2},y_{3})^{T}$ 
 and 
  $\vec{\pi}=-i(\frac{\partial}{\partial y_{1}},  \frac{\partial}{\partial y_{2}},\frac{\partial}{\partial y_{3}})^{T}$ are their canonical momenta
 because  $ \mathbb{R} $ is orthogonal ($ \mathbb{R}^{-1}=\mathbb{R}^{T}$).
 The coupling matrix $\mathbb{C}$ is given by
 \begin{align}
 \mathbb{C}(t)=\text{diag}\left(\omega_{1}^{2},\omega_{2}^{2},\omega_{3}^{2}\right)+\tilde{\mathbb{C}}
 \end{align}
 such that  $\tilde{\mathbb{C}}_{ij}=\tilde{\mathbb{C}}_{ji}=-C_{ij}$, with $i<j$.
Now, our problem is reduced to find Euler eigenangles $(\theta,\psi,\phi)$ and  then we start  looking for the eigenvalues of  $\mathbb{C}(t)$. The resolution of the characteristic equation 
\begin{equation}
\sum\limits_{j=0}^{3}b_{j}(t)\lambda_{j}^{j}(t)=0, \qquad b_{3}(t)=1
\end{equation}
associated to $\mathbb{C}(t)$ gives rise to the set of eigenvalues $\sigma^{2}_{i}(t)$ 
\begin{eqnarray}
\sigma_{1}^{2}(t)&=&\frac{1}{3}\left[b_{2}(t)+2\sqrt{p(t)}\cos(\Phi(t))\right]\label{sig1}\\
\sigma_{2}^{2}(t)&=&\frac{1}{3}\left[b_{2}(t)+2\sqrt{p(t)}\cos\left(\Phi(t)+\frac{2\pi}{3}\right)\right]\label{sig2}\\
\sigma_{3}^{2}(t)&=&\frac{1}{3}\left[b_{2}(t)+2\sqrt{p(t)}\cos\left(\Phi(t)-\frac{2\pi}{3}
\right)\right]\label{sig3}
\end{eqnarray} 
where the time-dependent parameters are
\begin{eqnarray}
b_{0}(t)&=&\sum\limits_{(i,i)\neq(j,k),j<k}^{3}\omega^{2}_{i}(t)C_{jk}^{2}(t)-\prod\limits_{i=1}^{3}\omega^{2}_{i}(t)-2\prod\limits_{i<j}^{3}C_{ij}(t)\\
b_{1}(t)&=&\sum\limits_{i<j}^{3} \omega_{i}^{2}(t)\omega_{j}^{2}(t)-\sum\limits_{i<j}^{3}C_{ij}^{2}(t)\\
 b_{2}(t)&=&\sum\limits_{i=1}^{3}\omega^{2}_{i}(t), \qquad
 \Phi(t) =\frac{1}{3} \arctan\left(\frac{\sqrt{p^{3}(t)-q^{2}(t)}}{q(t)}\right) \\
p(t)&=& b_{2}^{2}(t)-3b_{1}(t), \qquad
q(t) =-\frac{27}{2}b_{0}(t)-b_{2}^{3}(t)+\frac{9}{2} b_{1}(t)b_{2}(t)
\end{eqnarray}
At this level, 
we have some comments in order. Firstly,
we show the relation
$\sum\limits_{j=1}^{3}\sigma_{j}^{2}(t)=\sum\limits_{j=1}^{3}\omega_{j}^{2}(t)$ between 
the eigenvalues of $\mathbb{C}(t)$ and frequencies. Secondly, the symmetry of the original frequencies (i.e. $ \omega_{k}=\omega_{j}$ $\forall j,$ $k\in\lbrace 1,2,3\rbrace $) 
does not entail that one of the normal frequencies (i.e. $ \sigma_{k}=\sigma_{j}$ $ \forall j,$ $k\in\lbrace 1,2,3\rbrace $).  Thirdly, 
  if the  energy spectrum is strictly positive then  we require a matrix $\mathbb{C}(t)$  positive. Consequently, by using the \textit{Sylvester criterion} we easily check that the coupling parameters and  frequencies should verify
  the  inequality 
 \begin{equation}
{\max(\omega_{1}^{2},0)\times \max(\omega_{2}^{2}\omega_{3}^{2}-J_{23}^{2},0)\times \max(\sigma_{1}^{2}\sigma_{2}^{2}\sigma_{3}^{2},0)>0}
 \end{equation}
    With the help of the \textit{identity eigenvalues-eigenvectors} theorem \cite{R1}, we  find the inputs of the Euler rotation $\mathbb{R}_{ij}$ 
	\begin{eqnarray}
 \mathbb{R}_{ij}^{2}(t)&=& \frac{\prod\limits_{k=1}^{2}\left(\sigma_{i}^{2}-\lambda_{k}(\mathbb{M}_{j})\right)}{\prod\limits_{k=1,k\neq i}^{3}{\left(\sigma_{i}^{2}-\sigma_{k}^{2}\right)}},\qquad \sum\limits_{j}^{3}\mathbb{R}_{ij}^{2}=1 
\end{eqnarray}
$\forall i\in\left\lbrace 1,3 \right\rbrace$
where the matrices $\mathbb{M}_{j}$ are the  minors obtained by simplifying the $j^{th}$ row and $j^{th}$ column of  $\mathbb{C}(t)$ and $\lambda_{k}$ are its eigenvalues. Then
 from ({\ref{E}}), we show that the Euler eigenangles can be expressed as
\begin{align}
&
\tan^{2}(\psi(t))=\frac{\mathbb{R}_{23}^{2}(t)}{\mathbb{R}_{13}^{2}(t)}=\frac{\left[\sigma_{2}^{4}(t)-(\omega_{1}^{2}(t)+\omega_{2}^{2}(t))\sigma_{2}^{2}(t)+\omega_{1}^{2}(t)\omega_{2}^{2}(t)-C_{12}^{2}(t)\right]\left(\sigma_{1}^{2}(t)-\sigma_{3}^{2}(t)\right)}{\left[\sigma_{1}^{4}(t)-(\omega_{1}^{2}(t)+\omega_{2}^{2}(t))\sigma_{1}^{2}(t)+\omega_{1}^{2}(t)\omega_{2}^{2}(t)-C_{12}^{2}(t)\right]\left(\sigma_{3}^{2}(t)-\sigma_{2}^{2}(t)\right)}\\ 
&
\cos^{2}(\theta(t))=\mathbb{R}^{2}_{33}(t)=\frac{\sigma_{3}^{4}(t)-(\omega_{1}^{2}(t)+\omega_{2}^{2}(t))\sigma_{3}^{2}(t)+\omega_{2}^{2}(t)\omega_{1}^{2}(t)-C_{12}^{2}(t)}{(\sigma_{3}^{2}(t)-\sigma_{1}^{2}(t))(\sigma_{3}^{2}(t)-\sigma_{2}^{2}(t))}\\
&
 \tan^{2}(\phi(t))=\frac{\mathbb{R}_{32}^{2}(t)}{\mathbb{R}_{31}^{2}(t)}= \frac{\sigma_{3}^{4}(t)-(\omega_{1}^{2}(t)+\omega_{3}^{2}(t))\sigma_{3}^{2}(t)+\omega_{1}^{2}(t)\omega_{3}^{2}(t)-C_{13}^{2}(t)}{\sigma_{3}^{4}(t)-(\omega_{2}^{2}(t)+\omega_{3}^{2}(t))\sigma_{3}^{2}(t)+\omega_{2}^{2}(t)\omega_{3}^{2}(t)-C_{23}^{2}(t)}.
\end{align} 
After this algebraic analysis, $H'$ 
takes the form 
\begin{equation}
H^{'}(y_{1},y_{2},y_{3},t)=\dfrac{1}{2}\vec{\pi}^{T}\vec{\pi}+\frac{1}{2}\vec{y}^{T}\mathbb{D}(t)\vec{y}+\sum_{i=1}^{3}a_{i}(t)\hat{L}_{i}.\label{hd}
\end{equation}As  clearly seen, it is still complicated to extract the solutions of energy spectrum  by directly solving the eigenvalue equation associated to \eqref{hd}. To overcome such situation, we proceed by adopting an interesting  model 
used in the literature.

\section[sqm]{Sudden quenched model \label{Sec3}}

   To decouple the Hamiltonian \eqref{hd}, we confine ourselves in the frame of sudden quenched model (SQM), in which the physical parameters are abruptly changed. This model appears at first non-physical, because of the discontinuity of physical parameters, but {such kind of time variation can be found for instance in a  $ LC $ circuit whose capacitor is  pumped by a voltage $ V(t) $ \cite{Richard}, or in a charged pendulum in alternating,
   	piece-wise constant, homogeneous electric field \cite{Burov}}. Besides, the eigen-energy of the system is time-independent  as originally showed by Lewis and Reisenfeld \cite{R29}. The dynamics of covariance matrix (CM), which encodes the information content of our quantum system,  is totally  governed by the solutions of the Ermakov equations (\ref{erm}). Consequently, if the dynamics of the Ermakov solutions in the framework of SQM  is nicely similar to those of the continuous model, then the use of SQM is 
  theoretically legitimate. It is the case for instance in \cite{R41}
  by  showing that  SQM  and the exponential behaviour are similar.  In addition,
  SQM is used to follow the dynamics of vacuum entanglement, mixing and quantum fulctuations in 3D- \cite{R4} and 2D- \cite{R5,R3,R21} coupled bosonic harmonic oscillators.
 Note that, SQM is necessary but not sufficient to remove the angular momenta term in (\ref{hd}).

  Motivated by the mentioned studies above
  and  to achieve our goal,  we consider a linear sudden quenched model (LSQM) for the coupling parameters $C_{ij}$ and   frequencies $\omega_{i}$, such as
\begin{equation}
\omega_{j}(t)=\left\{
\begin{array}{ll}
\omega_{j}(0) & t=0\\
\epsilon \omega_{j}(0) & 0<t\\
\end{array}
\right.,\qquad C_{kl}(t)=\left\{
\begin{array}{ll}
C_{kl}(0)& t=0\\
\epsilon^2 C_{kl}(0)& 0<t\\
\end{array}
\right.
\end{equation}
where 
 $\epsilon$ is a  dimensionless parameter has a paramount role in the next analysis, because it will promote the quantification of the quench and its effect on the dynamics, which will be called quench factor. Moreover, $\epsilon$ is a feasible parameter to engineer the optimality of  entanglement  and coherence  resources in a given time scale. It is interesting to note that when we set $\epsilon={\omega_{j}(t>0)}/{\omega_{0}}$, one can give  a suitable physical meaning of it. Indeed, it can be seen if the initial and final states are canonical as the rate of temperature decreasing in the frame of atom cooling in time-dependent harmonic traps \cite{R35}. For example the experiment of cooling with harmonic trap was realized by taking  the value  $\epsilon=0.1$ \cite{R42}. In general
  $ \epsilon \in \left]-\infty,+\infty \right[\setminus \left\lbrace 0 \right\rbrace $ but with the symmetry criterion of the Hamiltonian   $ H(\epsilon)=H(-\epsilon)$
  we have  $ \epsilon>0 $. It is worthy to note that in next analysis we will confine ourselves in the cooling regime, i.e.  $\epsilon\in\left]0,1\right]$.

 By applying LSQM, the rotation matrix  $\mathbb{R}$ becomes time-independent and then the Euler eigenangles ($\theta, \phi, \psi $) are now constant for all time, i.e.   $\dot{\theta}=\dot{\psi}=\dot{\phi}=0$.  Consequently, the Euler velocities $ a_{i}(t)=0, \, \forall t\geq 0 $, and  the last term in \eqref{hd} will be discarded $\sum\limits_{i=1}^{3}a_{i}(t)\hat{L}_{i}=0$. Consequently, we can easily derive  the solutions of energy spectrum of  $H$ from those of $H^{'}$ 
   \begin{eqnarray}
\Psi_{n,m,l}(y_{1},y_{2},y_{3}:t)&=&\mathcal{N}(t)\ e^{-i \left(f_{1,n}(t)+f_{2,m}(t)+f_{3,l}(t)\right)}\ e^{-\frac{1}{2}\sum\limits_{j=1}^{3}\left(\Omega_{j}(t)-i\frac{\dot{\rho}_{j}}{\rho_{j}}\right)y_{j}^{2}}\\\nonumber
&&\times H_{n}(\sqrt{\Omega_{1}(t)}y_{1}) H_{m}(\sqrt{\Omega_{2}(t)}y_{2}) H_{l}(\sqrt{\Omega_{3}(t)}y_{3})\nonumber
\end{eqnarray}
such that $H_{j}(\epsilon)$ are  Hermite functions and we have defined
the  time-dependent functions  
\begin{eqnarray}
\mathcal{N}(t)&=& 
\frac{\left(\prod\limits_{i=1}^{3}\Omega_{i}(t)\right)^{\frac{1}{4}}} 
{\sqrt{\pi^{3/2} 2^{n+m+l}n!m!l!}}\\
f_{i,j}(t)&=& \left(j+\frac{1}{2}\right)\int_{0}^{t}\Omega_{i}(s)ds, \qquad \Omega_{i}(s)=\frac{\sigma_{i}(0)}{\rho_{i}^{2}(s)}
\end{eqnarray}
where  $i=1,2,3$,  $j=n,m,l$ and 
the scale factors $\rho_{i}(t)$ verify the Ermakov non-linear equation \cite{R6}
\begin{equation}
\ddot{\rho}_{i}+\sigma_{i}^{2}(t)\rho_{i}(t)=\frac{\sigma_{i}^{2}(0)}{\rho_{i}^{3}(t)}
\label{erm}
\end{equation}
with the conditions $\rho_{i}(0)=1$ and $\dot{\rho_{i}}(0)=0$ to guaranty  the unitarity of dilatation operator used for each single time-dependent harmonic oscillator \cite{R29,R6}.  The solutions of \eqref{erm} in the framework of LSQM are
\begin{equation}
{\rho_{i}(t)=\frac{1}{\sqrt{2}\epsilon}\sqrt{(\epsilon^{2}-1)\cos\left(2\epsilon \sigma_{i}(0)t \right)+\epsilon^{2}+1}} 
\end{equation}where  $\sigma_{i}(0)$ are the initial normal frequencies of  modes $i=1,2,3$. Note that the difference between  Ermakov modes is $\sigma_{i}(0)$ but the amplitude is the same for  three modes $ \rho_{i}(t) $. Now by performing the rotation rule  on  $\Psi_{n,m,l}(y_{1},y_{2},y_{3};t)$ \cite{R6}, we get the eigenfunctions of $H$
\begin{eqnarray}
\Psi_{n,m,l}(x_{1},x_{2},x_{3};t)&=&\mathcal{N}(t)\ e^{-i \left(f_{1,n}(t)+f_{2,m}(t)+f_{3,l}(t)\right)} e^{-\frac{1}{2}\sum\limits_{i=1}^3\left(\Omega_{i}(t)-i\frac{\dot{\rho}_{i}}{\rho_{i}}\right)\left(\sum\limits _{j=1}^{3}\mathbb{R}_{ij}x_{j}\right)^2}\nonumber\\
&&\times \prod_{k=n,m,l;i=1}^{3}H_{k}\left(\sqrt{\Omega_{i}(t)}\sum \limits_{j=1}^{3}\mathbb{R}_{ij}x_{j}\right).
\end{eqnarray}
In the forthcoming analysis, we will be only 
interested to the vacuum solution $\Psi_{0}(x_{1},x_{2},x_{3};t)$ associated  to  the density matrix 
\begin{equation}
 \rho_{ABC}(x_{1},x_{2},x_{3},z_{1},z_{2},z_{3};t)=\sqrt{\frac{\prod\limits_{i=1}^{3}\Omega_{i}(t)}{\pi^{3}}}e^{-\frac{1}{2}\sum \limits_{i,j=1}^3\left( x_i A_{ij}(t) x_j +z_{i} A_{ij}^{\ast}(t) z_j\right)}
\end{equation}
 where the matrix $A$ is time-dependent and symmetric $(A_{ij}=A_{ji})$ 
with the  elements
 \begin{eqnarray}
 A_{jj}=\sum\limits_{i=1}^3 \overline{\Omega}_{i}(t)\mathbb{R}_{ij}^{2},\qquad 
 A_{ij}=\sum\limits_{k=1}^{3}\overline{\Omega}_{k}(t)\mathbb{R}_{ki}\mathbb{R}_{kj}
 \end{eqnarray}
 and the complex frequencies read as 
 \begin{equation}
 \overline{\Omega}_{i}(t)=\Omega_{i}(t)-i\frac{\dot{\rho}_{i}}{\rho_{i}}(t).
 \end{equation}
 {These obtained results will be employed  in the forthcoming analysis  to investigate different physical quantities.}
 
 \section{Wigner distribution and covariance matrix\label{sect2}}
 
 To study the quantum fluctuations for our system, we  consider
the {Wigner} distribution 
associated to vacuum state  
\begin{eqnarray}
\mathcal{W}_{0}(x_{1},x_{2},x_{3}:p_{1},p_{2},p_{3}:t)&:=&\frac{1}{\pi^{3}}\int dq_{1}dq_{2}dq_{3}\Psi_{0}^{\ast}\left(x_{1}+q_{1},x_{2}+q_{2},x_{3}+q_{3}:t\right)\\
&&\times  \Psi_{0}\left(x_{1}-q_{1},x_{2}-q_{2},x_{3}-q_{3}\right)e^{-2i(p_{1}q_{1}+p_{2}q_{2}+p_{3}q_{3})}. \nonumber 
\end{eqnarray}
However, the computation of such integral is very tedious, then we use the fundamental property of $\mathcal{W}_{0}(x_{1},p_{1},x_{2},p_{2},x_{3},p_{3})$. Indeed,  let $U(R)$  an infinite dimensional  unitary  operator  corresponding  to $R \in Sp(6,\mathbb{R})$, which transforms the state  $\Psi$ 
 to $\Psi^{'}=U(R)\Psi$. It follows that  the density matrix and Wigner distribution will be changed as \cite{R7}
\begin{eqnarray}
 \rho'_{ABC}=U(R)\rho_{ABC} U(R)^{-1},\qquad 
 \mathcal{W}_{\rho'_{ABC}}(\xi ;t)=\mathcal{W}_{\rho_{ABC}}(R^{-1}\xi;t)                                 
\label{w}\end{eqnarray}
and  $\xi=\left(x_{1},p_{1},x_{2},p_{2},x_{3},p_{3}\right)^{T}$ is the phase space vector. By using  (\ref{w}), we show
\begin{eqnarray}
\mathcal{W}_{\rho^{'}}(x_{1},x_{2},x_{3},p_{1},p_{2},p_{3};t)=\frac{1}{\pi^{3}}e ^{\left[-\xi^{T}\mathbb{S}(t)\xi\right]}
\end{eqnarray}
such that the matrix $\mathbb{S}(t)$ is symmetric, real and semi-definite positive 
\begin{eqnarray}
\mathbb{S}(t)=
  \begin{pmatrix}
   \mathbb{S}_{1}(t)& \mathbb{O}_{2\times 2} & \mathbb{O}_{2\times 2} \\
    \mathbb{O}_{2\times 2} &   \mathbb{S}_{2}(t)& \mathbb{O}_{2\times 2} \\
    \mathbb{O}_{2\times 2} & \mathbb{O}_{2\times 2} &   \mathbb{S}_{3}(t) \\
  \end{pmatrix},\qquad
\mathbb{S}_{i}(t)= \begin{pmatrix}  A_{i}(t)& C_{i}(t) \\
                       C_{i}(t) &   B_{i}(t)
                       \end{pmatrix}
                         \end{eqnarray}
where  different elements read as 
\begin{eqnarray}
A_{i}(t)= \Omega_{i}(t)+\frac{1}{\Omega_{i}(t)}\left(\frac{\dot{\rho_{i}}(t)}{\rho_{i}(t)}\right)^{2}, \qquad
B_{i}(t)=\frac{1}{\Omega_{i}(t)}, \qquad 
C_{i}(t)= \frac{1}{\Omega_{i}(t)}\frac{\dot{\rho_{i}}(t)}{\rho_{i}(t)}\label{eq39}
\end{eqnarray}
and $\mathbb{O}_{2\times 2}$ is $2\times 2$ zero matrix,  
 with  $i=1,2,3$. From transformation inverse, we get 
\begin{eqnarray}
\mathcal{W}_{\rho}(x_{1},p_{1},x_{2},p_{2},x_{3},p_{3}:t)&=&\frac{1}{\pi^{3}}e^{\left[-\xi^{T}\mathbb{G}(t)\xi\right]} \label{g}
\end{eqnarray}
such that  the Gaussian matrix  $\mathbb{G}(t)=R^{T}\mathbb{S}R$,
 $2 \times 2$ block  $R_{ij}=\bigoplus\limits_{i=1}^{2}\mathbb{R}_{ij}$,
 has the  elements 
\begin{eqnarray}
\mathbb{G}_{11}(t)&=&\sum\limits_{i=1}^{3}A_{i}(t)\mathbb{R}_{i1}^{2}(t), \quad G_{33}(t)=\sum\limits_{i=1}^{3}A_{i}(t)\mathbb{R}_{i2}^{2}(t),\quad \mathbb{G}_{55}(t)=\sum\limits_{i=1}^{3}A_{i}(t)\mathbb{R}_{i3}^{2}(t) \nonumber\\
\mathbb{G}_{22}(t)&=&\sum\limits_{i=1}^{3}B_{i}(t)\mathbb{R}_{i1}^{2}(t),\quad \mathbb{G}_{44}(t)=\sum\limits_{i=1}^{3}B_{i}(t)\mathbb{R}_{i2}^{2}(t),\quad \mathbb{G}_{66}(t)=\sum\limits_{i=1}^{3}B_{i}(t)\mathbb{R}_{i3}^{2}(t)\nonumber\\
\mathbb{G}_{13}(t)&=&\sum\limits_{i=1}^{3}A_{i}(t)\mathbb{R}_{i1}(t)\mathbb{R}_{i2}(t),
\quad  \mathbb{G}_{15}(t)=\sum\limits_{i=1}^{3}A_{i}(t)\mathbb{R}_{i1}(t)\mathbb{R}_{i3}(t),
\quad
\mathbb{G}_{35}(t)= \sum\limits_{i=1}^{3}A_{i}(t)\mathbb{R}_{i2}(t)\mathbb{R}_{i3}(t)\nonumber \\                  
\mathbb{G}_{24}(t) &=& \sum\limits_{i=1}^{3}B_{i}(t)\mathbb{R}_{i1}(t)\mathbb{R}_{i2}(t), \quad  
\mathbb{G}_{26}(t)=\sum\limits_{i=1}^{3}B_{i}(t)\mathbb{R}_{i1}(t)\mathbb{R}_{i3}(t),\quad \mathbb{G}_{46}(t)=\sum\limits_{i=1}^{3}B_{i}(t)\mathbb{R}_{i2}(t)\mathbb{R}_{i3}(t) \nonumber\\
\mathbb{G}_{12}(t)&=&\sum\limits_{i=1}^{3}C_{i}(t)\mathbb{R}_{i1}^{2}(t),\quad \mathbb{G}_{34}(t)=\sum\limits_{i=1}^{3}C_{i}(t)\mathbb{R}_{i2}^{2}(t),\quad \mathbb{G}_{56}(t)=\sum\limits_{i=1}^{3}C_{i}(t)\mathbb{R}_{i3}^{2}(t)\nonumber\\
\mathbb{G}_{14}(t)&=&\sum\limits_{i=1}^{3}C_{i}(t)\mathbb{R}_{i1}(t)\mathbb{R}_{i2}(t),\quad\mathbb{G}_{16}(t)=\sum\limits_{i=1}^{3}C_{i}(t)\mathbb{R}_{i1}(t)\mathbb{R}_{i3}(t), \quad  
\mathbb{G}_{36}(t)=\sum\limits_{i=1}^{3}C_{i}(t)\mathbb{R}_{i2}(t)\mathbb{R}_{i3}(t)
\nonumber
\end{eqnarray} 
with $\mathbb{G}_{14}=\mathbb{G}_{23}$, $\mathbb{G}_{16}=\mathbb{G}_{25}$
and $\mathbb{G}_{36}=\mathbb{G}_{45}$.
Note that, $\mathbb{G}(t)$ is  positive semi-definite, real and symmetric with $\det\mathbb{G}(t)=1$ (the global state is pure). In our case the state is Gaussian then we have  $\mathbb{G}(t)=\sigma^{-1}(t)$ such that $\sigma(t)$ is the covariance matrix, which encodes all the quantum information of our state. To explicitly give $\sigma(t)$, we recall that the quadrature (phase space) vector is defined by 
\begin{equation}
\hat{Q}= \left(\hat{q}_{1},\hat{p}_{1}:\hat{q}_{2},\hat{p}_{2}:\hat{q}_{3},\hat{p}_{3}\right)^{T},\qquad \left[\hat{Q}_{k},\hat{Q}_{l}\right]=i\Omega_{kl},\qquad k,l=1,\cdots,6
\end{equation}
where $\Omega$ is the symplectic form matrix 
\begin{equation}
\Omega=\bigoplus_{i=1}^{3}
  \begin{pmatrix}
    0&1\\
    -1&0 \\
  \end{pmatrix}
\end{equation} 
and use 
{\sc Theorem}: {\it Let $\mathbb{G}(t)$ and $\sigma(t)$  be the Gaussian matrix of the Wigner distribution and covariance matrix, respectively. If the state $\sigma(t)$ is pure then} we have
\begin{equation}
{\sigma}(t)=\Omega^{-1}\mathbb{G}(t)\Omega.
\end{equation}
$\blacksquare$ {\sc Proof}: {\it If $\sigma(t)$ is pure state  then  under a symplectic transformation $S$ 
	it will be similar to unit matrix $\mathbb{I}_{6}$ (Williamson theorem \cite{R9}),
	i.e.  $\sigma(t)= S\mathbb{I}_{6}S^{T} $ and $S\Omega S^{T}=\Omega$. In addition, we have  $\sigma(t)^{-1}=\mathbb{G}(t)$, then it suffices to show the relation  $\mathbb{G}(t)\Omega^{-1}\mathbb{G}(t)\Omega=\mathbb{I}_{6}$. Since  $\mathbb{G}^{-1}=(SS^{T})^{-1}$, it follows that } 
	\begin{equation}
	\mathbb{G}(t)\Omega^{-1}\mathbb{G}(t)\Omega=(SS^{T})^{-1}\Omega^{-1}(SS^{T})^{-1}\Omega=\left(\Omega^{-1}SS^{T}\Omega SS^{T}\right)^{-1}=\mathbb{I}_{6} \ \blacksquare
\end{equation}
One can also use the relation $\mathbb{G}(t)=R^{T}\mathbb{S}R$ to obtain  $\sigma(t)=R^{T}\mathbb{S}^{-1}(t)R$ with $\det\mathbb{S}_{i}=1$ and
\begin{eqnarray}
\mathbb{S}^{-1}(t)=\bigoplus\limits_{i=1}^{3}\tilde{\mathbb{S}}_{i}(t), \qquad  \tilde{\mathbb{S}}_{i}(t)=
  \begin{pmatrix}
    B_{i}(t) &-C_{i}(t) \\
    -C_{i}(t) & A_{i}(t) \\
  \end{pmatrix}. 
  \end{eqnarray}
  Finally,  we  end up with the covariance matrix 
\begin{eqnarray}
\sigma(t)= \begin{pmatrix} \mathbb{G}_{{22}}&-\mathbb{G}_{{12}}&\mathbb{G}_{{24}}&-\mathbb{G}_{{23}}&\mathbb{G}
_{{26}}&-\mathbb{G}_{{25}}\\ \noalign{\medskip}-\mathbb{G}_{{12}}&\mathbb{G}_{{11}}&-\mathbb{G}_{{23}}&\mathbb{G}_{
{13}}&-\mathbb{G}_{{25}}&\mathbb{G}_{{15}}\\ \noalign{\medskip}\mathbb{G}_{{24}}&-\mathbb{G}_{{23}}&\mathbb{G}_{{44
}}&-\mathbb{G}_{{34}}&\mathbb{G}_{{46}}&-\mathbb{G}_{{36}}\\ \noalign{\medskip}-\mathbb{G}_{{23}}&\mathbb{G}_{{13}}
&-\mathbb{G}_{{34}}&\mathbb{G}_{{33}}&-\mathbb{G}_{{36}}&\mathbb{G}_{{35}}\\ \noalign{\medskip}\mathbb{G}_{{26}}&-\mathbb{G}
_{{25}}&\mathbb{G}_{{46}}&-\mathbb{G}_{{36}}&\mathbb{G}_{{66}}&-\mathbb{G}_{{56}}\\ \noalign{\medskip}-\mathbb{G}_{
{25}}&\mathbb{G}_{{15}}&-\mathbb{G}_{{36}}&\mathbb{G}_{{35}}&-\mathbb{G}_{{56}}&\mathbb{G}_{{55}}\end{pmatrix}
 =\begin{pmatrix}
\sigma_{A}&\Upsilon_{A,B}&\Upsilon_{A,C}\\
\Upsilon_{A,B}^{T}&\sigma_{B}&\Upsilon_{B,C}\\
\Upsilon_{A,C}^{T}&\Upsilon_{B,C}^{T}&\sigma_{C}
\end{pmatrix}
 \label{si}\end{eqnarray}
where the block matrix $\sigma_{m}$ is the local covariance matrix corresponding to the marginal (reduced state) of mode $m$, and the off-diagonal matrices $\Upsilon_{pq}$ are the inter-modal correlations between the modes $p$ and $q$, which vanish for  a separable state $\sigma_{ABC}=\sigma_{A}\oplus\sigma_{B}\oplus\sigma_{c}$. 
It is worthy to emphasis that  the inequality  $\sigma+i\Omega\geq0$ is  a necessary and sufficient condition  that must be fulfilled by $\sigma(t)$ in order to describe a physical  density matrix $\rho$ \cite{R7,R8}. Note that from (\ref{g}), the state is centered $\langle \hat{Q}_{i}\rangle=\Tr\left[\rho \hat{Q}_{i} \right]=0$ and then $\sigma(t)=\left(\sigma\right)_{ij}=\langle \lbrace\hat{Q}_{i},\hat{Q}_{j}\rbrace \rangle$ with $\lbrace, \rbrace$ stands for anti-commutator.

 \section{Dynamics of mixedness $ S_{L} $ and entropy of entanglement \label{sect3}}
 \subsection{Physicality and classification of pure Gaussian state} 

 It is known that the quantum mechanics is linear, which entails the no-cloning theorem,
 and  then the entanglement resource, quantified through the reduced  von Neumann entropy, must be monogamous.
 The fact that entanglement can not be freely shareable, at striking variance with the behavior of classical correlations,
 the local purities $P_{A,B,C}(t)=\mathfrak{a}^{-1}_{A,B,C}(t)$ should verify the following triangular inequality \cite{R11}   
 \begin{equation}
 \vert \mathfrak{a}_{i}(t)-\mathfrak{a}_{j}(t)\vert+1 \leq \mathfrak{a}_{k}(t)\leq \mathfrak{a}_{i}(t)+\mathfrak{a}_{j}(t)-1 
 \label{Strongcondition}
 \end{equation}
 which is invariant under permutation of labels $i,j,k\in\left\lbrace A,B,C \right\rbrace $. It  
  is sufficient and necessary  to guaranty the physicality of 
  $\sigma(t)$  (\ref{si}). Now the main question  how to  choose the parameters $C_{ij}(t)$ and $\omega_{i}(t)$ in order to preserve such strong 
  condition during the dynamics.
 To give an answer, we set two functions based on the local maxidness   
 $\lbrace{\mathfrak{a}_{i}}\rbrace$, let say 
 \begin{equation}
  S_{1}^{k}(t)=\mathfrak{a}_{k}-|\mathfrak{a}_{i}-\mathfrak{a}_{j}|-1, \qquad  S_{2}^{k}(t)=\mathfrak{a}_{i}+\mathfrak{a}_{j}-\mathfrak{a}_{k}-1
  \end{equation}
   Consequently, (\ref{Strongcondition}) is  equivalent to 
 \begin{eqnarray} 
 S_{1}^{k}(t)\geq 0,\qquad S_{2}^{k}(t)\geq 0.\label{ineq62}
 \end{eqnarray}
In the case of multipartite systems, there are several types of entanglement due to many ways by which  different subsystems may be entangled with each other. For a general Gaussian state $\rho$ (density matrix) {$\equiv \sigma$ (covariance matrix)}, we have  five  classes of states  \cite{R10}
\begin{itemize}
	\item $\mathcal{C}_{1}$: Fully inseparable states
	   \item $\mathcal{C}_{2}$: 1-mode biseparable states
	   \item $\mathcal{C}_{3}$: 2-mode biseparable states
	   \item $\mathcal{C}_{4}$: 3-mode biseparable states
	   \item $\mathcal{C}_{5}$: Fully separable  states.
\end{itemize} 
Note that $ \bigcup\limits_{j=1}^{5}\mathcal{C}_{j}=\mathcal{G} $ ($ \mathcal{G}:= $ the set of Gaussian states) and $\mathcal{C}_{j} \bigcap \mathcal{C}_{l}=\left\lbrace 0\right\rbrace $ with $1\leq j<l\leq 5$. In the case of pure state, the state does not belong to $\mathcal{C}_{4}$ and $\mathcal{C}_{3}$ \cite{R11}. The state will be fully inseparable when the the following condition is fulfilled for all modes \cite{R18}
\begin{eqnarray}
|P_{i}^{-1}(t,\epsilon)-P_{j}^{-1}(t,\epsilon)|+1<P^{-1}_{k}(t,\epsilon)<\sqrt{P^{-2}_{i}(t,\epsilon)+P^{-2}_{j}(t,\epsilon)-1}. \label{fs}
\end{eqnarray}

 \subsection{Mixedness and entanglement}
 
\hspace{3mm}The  characterization  of  the tripartite entanglement  of  three-mode $(ABC)$
Gaussian state is possible because the  positive partial transpose (PPT) criterion is necessary and sufficient for their separability under any, partial $(i\vert j )$ or global $(i\vert (jk))$ bipartition \cite{R10}.  Our state is a pure Gaussian state,  i.e $\det \mathbb{G}(t)=\det\sigma(t)=1$. The local purities $P_{j}(t)$ $(j=A,B,C$), suffice to quantify any quantum features (entanglement, correlations, $\cdots$) encoded in our state. To compute these purities, we use the partial tracing in the phase space $(\dim=6)$, rather than 
integrals in Hilbert space $(\dim=+\infty)$ \cite{R3,R4,R21}. In particular, the partial tracing is very handy to do at the phase space picture, because it reduces the whole state to each mode $m$, which can simply  be   achieved by just removing the block rows and columns pertaining to the excluded modes  $l$ $(l \neq m)$ \cite{R18}. 

Now by performing the partial tracing on our state, 
the reduced two modes $\sigma_{(AB)}(t), \sigma_{(BC)}(t), \sigma_{(AC)}$ and single mode states $\sigma_{A}(t),\sigma_{B}(t),\sigma_{C}(t)$ take the forms
\begin{eqnarray}
\sigma_{(AB)}(t)&=&
  \begin{pmatrix}
  \sigma_{A}(t) & \Upsilon_{A,B}(t) \\
    \Upsilon_{A,B}^{T}(t) & \sigma_{B}(t) \\
  \end{pmatrix},\qquad \sigma_{A}(t)= \begin{pmatrix}
  \mathbb{G}_{22}(t) & -\mathbb{G}_{12}(t)\\
    -\mathbb{G}_{11}(t) & \mathbb{G}_{11}(t) \\
   \end{pmatrix}\\ 
 \sigma_{(AC)}(t)&=& \begin{pmatrix}
  \sigma_{A}(t) & \Upsilon_{A,C}(t) \\
    \Upsilon_{A,C}^{T}(t) & \sigma_{C}(t) \\
   \end{pmatrix},\qquad
\sigma_{B}(t)= \begin{pmatrix}
  \mathbb{G}_{44}(t) & -\mathbb{G}_{34}(t)\\
    -\mathbb{G}_{34}(t) & \mathbb{G}_{33}(t) \\
   \end{pmatrix}\\
\sigma_{(BC)}(t)&=& \begin{pmatrix}
  \sigma_{B}(t) & \Upsilon_{B,C}(t) \\
    \Upsilon_{B,C}^{T}(t) & \sigma_{C}(t) \\
   \end{pmatrix},\qquad
\sigma_{C}(t)= \begin{pmatrix}
  \mathbb{G}_{66}(t) & -\mathbb{G}_{56}(t)\\
    -\mathbb{G}_{56}(t) & \mathbb{G}_{55}(t) \\
  \end{pmatrix}.
\end{eqnarray}
The purity of  mode $i$ is equal to those of modes $j$ and $k$ because the global state $(ijk)$ is pure. From the marginal  purities definitions \cite{R11}, one can show that for mode  $k$, the purity is 
\begin{eqnarray} \label{purk} 
P_{k}^{-2}(t,\epsilon)=1+\sum\limits_{m<j}^{3}\mathbb{R}_{mk}^{2}\mathbb{R}_{jk}^{2}\left[ \left( \sqrt{\frac{\sigma_{j}(0)}{\sigma_{m}(0)}}\dfrac{\rho_{m}}{\rho_{j}}-\sqrt{\frac{\sigma_{m}(0)}{\sigma_{j}(0)}}\dfrac{\rho_{j}}{\rho_{m}}\right)^{2}+\frac{1}{\sigma_{j}(0)\sigma_{m}(0)}\left( \rho_{m}\dot{\rho_{j}}-\rho_{j}\dot{\rho_{m}}\right)^{2}\right] 
\end{eqnarray}  
where $\rho\equiv \rho(t,\epsilon) $ and $\dot{\rho}\equiv \dot{\rho}(t,\epsilon) $. {The fact that the marginal purities depend on the modes $ (\rho_{1},\rho_{2},\rho_{3})$  solutions of Ermakov equations has an intriguing consequence,
	because one can show that such equations are unitary equivalent to the Hill ones \cite{Hill} for the classical
	counterpart of our quantum decoupled harmonic oscillators. Such amazing feature allows to study
	the optimal generation of entanglement only by designing the classical instabilities of the classical
	oscillators\cite{Roversi}. This will open an interesting gateway to understand the link between quantum sytems
	and their classical counterpart as well as orienting experiments to engineer the classical systems instead
	of their qunatum analogs, where the financial requirements are much important \cite{R32,R322} }.
Note that, 
for a time-independent Hamiltonian $(\epsilon=1)$, 
\eqref{purk} reduces to the following
\begin{eqnarray}
P_{k}^{-2}(t)=1+\sum\limits_{m<j}^{3}\mathbb{R}_{mk}^{2}\mathbb{R}_{jk}^{2}\left( \sqrt{\frac{\sigma_{m}(0)}{\sigma_{j}(0)}}-\sqrt{\frac{\sigma_{j}(0)}{\sigma_{m}(0)}}\right)^{2} \geq 1
\end{eqnarray}
since we have  $ \rho_{m}=\rho_{l}=1$ and $ \dot{\rho_{m}}=\dot{\rho_{l}}=0$.
 Another interesting case occurs when the three eigenvalues of the coupling matrix $\mathbb{C}(t) $ are equal giving rise to  purities equal  one, which means that the  state  belongs to  class 5 and then it  is fully separable. 
From these particular cases, the Ermakov modes do not play any role because they  vanish and then  $ \epsilon $ matters only when the Ermakov modes exist. This point will  help us to understand deeply the dynamics behavior of   entanglement, which is properly quantified using the von Neumann entropy $S_{v}$. Consequently  for each mode 
it reads as  
\begin{eqnarray}
S_{v}^{k}(t,\epsilon)=\frac{P_{k}^{-1}(t,\epsilon)+1}{2}\ln\left(\frac{P_{k}^{-1}(t,\epsilon)+1}{2}\right)-\frac{P_{k}^{-1}(t,\epsilon)-1}{2}\ln\left(\frac{P_{k}^{-1}(t,\epsilon)-1}{2}\right).
\end{eqnarray}
Note that if $P_{k} \longrightarrow 1$ (pure state) the state will be disentangled $S_{v}\longrightarrow 0$ but if $P_{k} \longrightarrow 0$ (maximally mixed state) $S_{v}$  diverges.
 
 \subsection{Fully symmetric state $\sigma_{A}=\sigma_{B}=\sigma_{C}$}
 
 The symmetric  Gaussian state has an invariant symmetric covariant matrix under the permutation of modes $A,B$ and $C$ \cite{R11}, i.e. the Hamiltonian is invariant under a permutation of the quadrature $(x_{i},p_{i})$ \cite{R32}. By  choosing the time-dependent coupling $C_{ij}(t)$ and frequencies $\omega_{k}(t)$ to be equal, then we can omit the labels $i,j$ and $k$ without losing of generality. In this case,
  the eigenvalues Eqs.(\ref{sig1}-\ref{sig3}) reduce to
  \begin{eqnarray}
   && \sigma_{1}^{2}(t) =\omega^{2}(t,\epsilon) +2C(t,\epsilon))\\ && \sigma_{2}^{2}(t,\epsilon)=\sigma_{3}^{2}(t,\epsilon)=
   \omega^{2}(t,\epsilon)-C(t,\epsilon)
  \end{eqnarray}
  and the rotation matrix  takes the  form
 \begin{eqnarray}
 \mathbb{R}= \begin{pmatrix}
\dfrac{1}{\sqrt{3}} & \dfrac{1}{\sqrt{3}} & \dfrac{1}{\sqrt{3}} \\ 
0 &\dfrac{1}{\sqrt{2}}  & -\dfrac{1}{\sqrt{2}}   \\ 
\sqrt{\frac{2}{3}} & \dfrac{1}{\sqrt{6}} & \dfrac{1}{\sqrt{6}}
  \end{pmatrix}.
 \end{eqnarray}  
Now we have  $ \Phi=0 $ and then the discriminant of the characteristic equation is  $ \Delta\propto (p^{3}-q^{2})=0 $. 
For simplicity,  
we choose  $ \omega^{2}(0)=2C(0)$ in the forthcoming analysis. Correspondingly, the Hamiltonian will be  reduced  to that of a closed Hooke chain,  i.e. the potential $ V(\vec{x},t)=\frac{C(t)}{2}\sum\limits_{j<k} ^{3}\left(x_{j}-x_{k}\right)^{2}$. The purity function becomes
\begin{eqnarray}  
P^{-2}(t,\epsilon)=1+\frac{2}{9}\left[2 \left( \dfrac{\rho_{2}}{\rho_{1}}-\dfrac{\rho_{1}}{2\rho_{2}}\right)^{2}+\frac{1}{2C(0)}\left( \rho_{2}\dot{\rho_{1}}-\rho_{1}\dot{\rho_{2}}\right)^{2}\right] .
\end{eqnarray}
We notice  that the necessary parameters  to describe the dynamics of  quantum correlation encoded in our Gaussian state are  $C(0)$ and $\epsilon$. To end the discussion on the physicality of the state, we show that the triangular inequality (\ref{ineq62}) will be reduced to the simple one $P(t,\epsilon)\leq1$.

 	\subsubsection{Dynamics of mixedness  and  effects of  $ C(0) $ and $\epsilon$}

The mixdness of a quantum state is a classical statistical feature, at variance  with quantum superposition that  is a purely quantum one. The difference between them
is that the second one 
 is a statistical feature encoding in the quantum state, but the first one is linked to the environment and classical distribution, which   does not have any relation with quantum structure of the state \cite{R21}. Physically speaking, the mixedness quantifies the lack of information about the preparation of the state. To quantify the amount of mixedness of our system, we use the linear entropy for continuous variable 
 \begin{equation}
 S_{L}(t,\epsilon)=1-P(t,\epsilon).
 \end{equation}
 
\begin{figure}[ht]
  	\centering
  	\includegraphics[width=6.4cm,height=5cm]{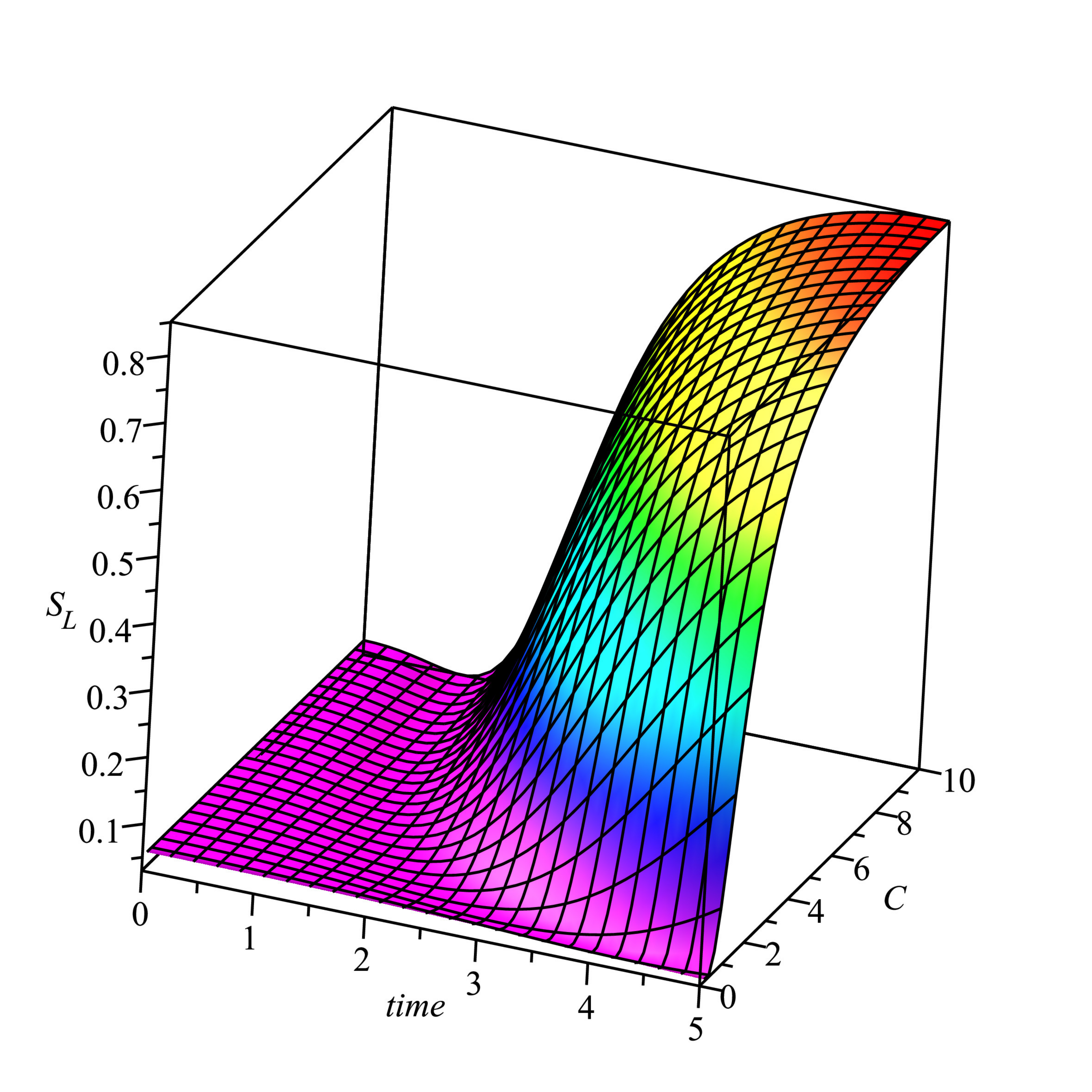}
  	\hspace{2cm}
  	\includegraphics[width=6.4cm, height=5cm]{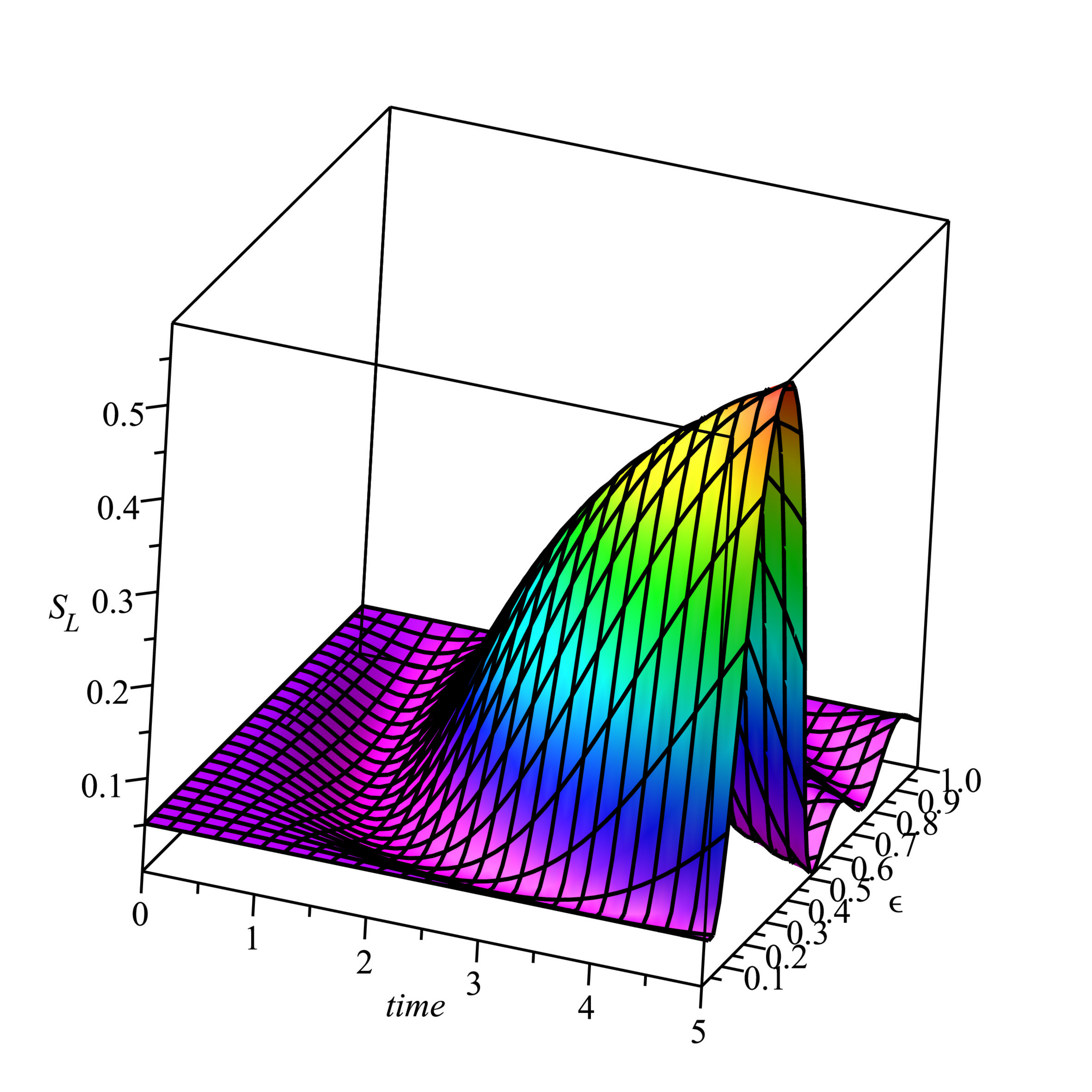}
  	\captionof{figure}{\sf (color online) The effects of
  		 the initial coupling $ C:=C(0) $ (left panel with $\epsilon=0.1$) 
  		 and the quench factor $\epsilon$ (right panel with C(0)=1) on the dynamics of mixedness in the time scale $[0,5]$. }\label{fig1}
  \end{figure}

  To follow the dynamics of mixendness of the physical state  and investigate the effect of the initial coupling $ C(0) $ and the quench factor $ \epsilon $ on the dynamics we present Figure \ref{fig1}. In left panel, we plot   $ S_{L} (t,\epsilon)$ versus   $ C(0) $ and the  time  scale  $t$ for a fixed value of the quench factor  $ \epsilon=0.1 $. The first observation from the dynamics  in the time scale $ t\in\left[ 0,5\right]  $ is that the creation of mixedness requires   certain time
  and  decreases as $ C(0) $  becomes important. This is due to the fact that for small values of $ C(0) $ the phase of the trigonometric function becomes also  small and then the frequency of Ermakov modes is $ \rho_{j}\sim 1 $.  The second one is that the extremal value of  $ S_{L}(t,\epsilon) $  is independent of $ C(0) $ because it modulates only 
  $ \rho_{j} $  and does not affect the maximal amount of mixedness. Consequently, we conclude that the key effect here is to control the frequency of oscillations.
 In right panel, 
 we investigate the effect of the quench factor $ \epsilon $ on
 the linear entropy $ S_{L}(t,\epsilon) $ for  the value $C(0) =1$. We observe that $ \epsilon $ contributes on the modulation of the amplitude together with the frequency of oscillations. Note that the degree of mixdness decreases as long as the quench factor increases and  the same effect on the frequency  of oscillations  with respect to the initial coupling $ C(0) $. It is intersting to stress that  the dynamics is governed by two modes  $\rho_{1}$ and $\rho_{2}=\rho_{3}$, which have  the same amplitude $ \frac{\epsilon^{2}-1}{2 \epsilon^{2}} $ but with a frequency hierarchy $\Delta\sigma= \epsilon\left|\sigma_{1}(0)-\sigma_{2}(0)\right|$. This latter
 is very important to create entanglement through the dynamics, otherwise we will have  $ S_{L}(t,\epsilon) =0$.

 \subsubsection{Dynamics of  entanglement and effect of $ C(0) $ and $ \epsilon $ }

\begin{figure}[ht]
 	\centering
 	\includegraphics[width=6.4cm, height=4cm]{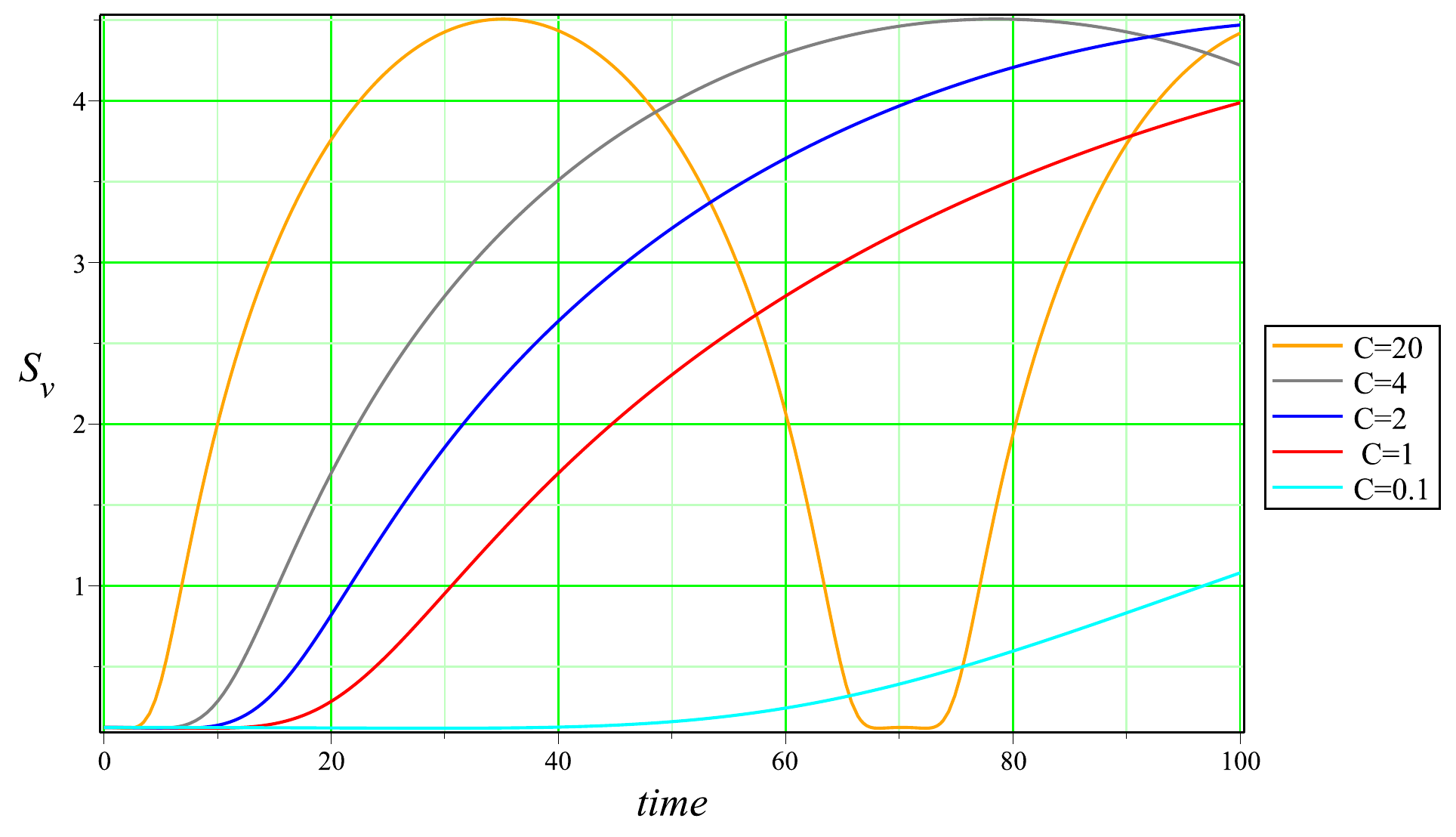}
 	\hspace{2cm}
 	\includegraphics[width=6.4cm, height=4cm]{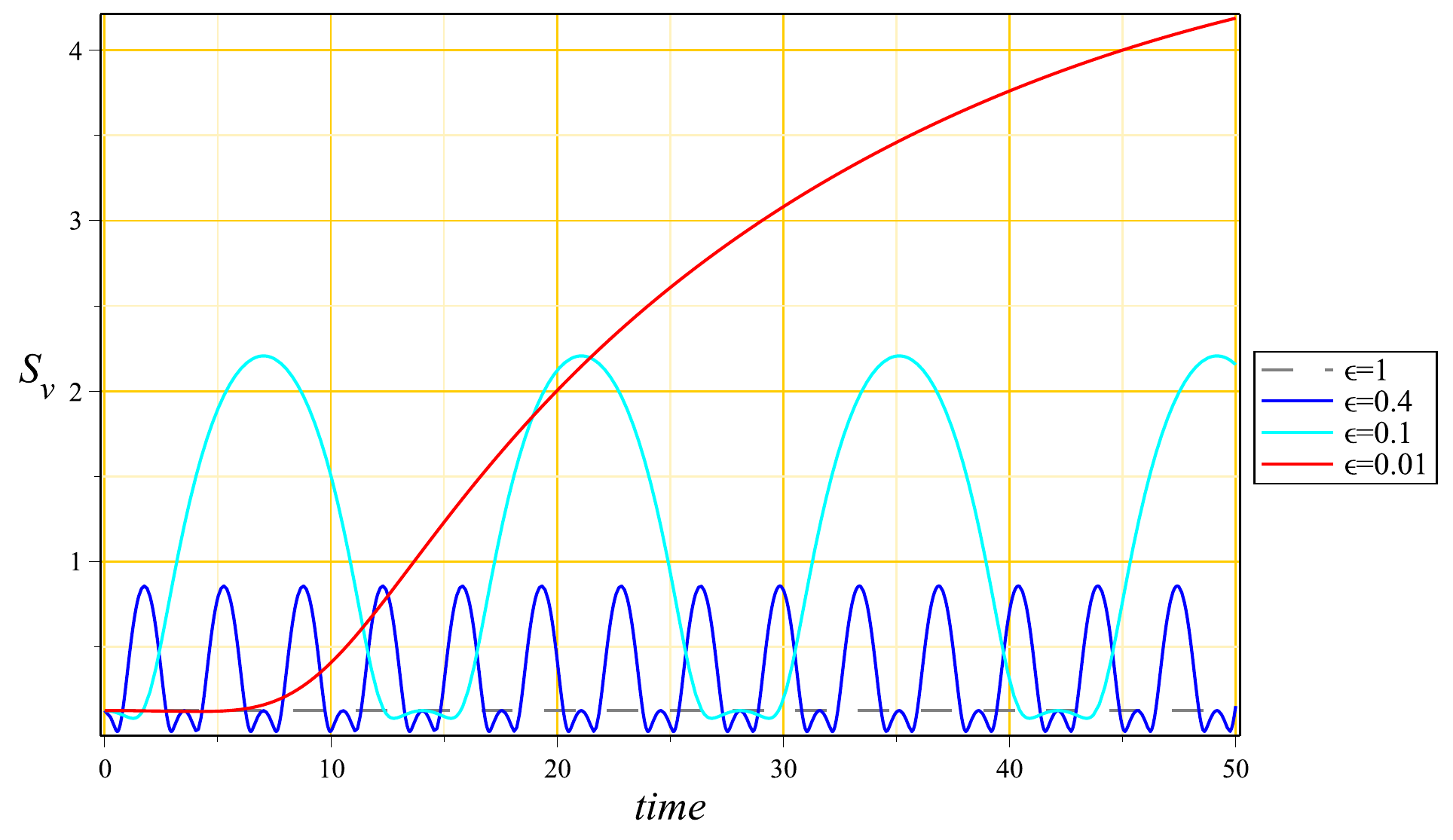}
 	\captionof{figure}{\sf (color online)
 		The effects of the initial coupling $ C(0):=C $   (left panel with $ \epsilon=0.1 $)
 		and  the quench factor $ \epsilon $ (right panel with $ C(0)=5 $)  on the the dynamics of entanglement  in the time scale $[0,100]$.}\label{fig2}
 \end{figure}

 To follow the dynamics of entanglement and the effects of the initial  coupling $ C(0) $ and the quench factor $\epsilon$, in Figure \ref{fig2} we plot  the dynamics  of von Neumann entropy $ S_{v} $ under some particular  values of  $ C(0) $ (left panel) and $\epsilon$ (right panel) in the time scale $ \left[ 0,100\right] $. 
  In left panel with $ \epsilon=0.1$, we observe that the dynamics requires a certain time to establish the entanglement,  which is due to  the phase of the Ermakov modes $ \sim \epsilon \sqrt{C(0)} t $ and then to engineer the time one can change $ C(0) $ or $ \epsilon $ or both. The amount of entanglement is independent of $ C(0) $ and  modulates only the frequency of modulation, because in the expression of $ S_{v} $ the $ C(0)  $ parameter is always included in the phase of the trigonometric functions ($ \cos $ and $ \sin $).
  In right panel,  we  set the coupling to $ C(0)=5 $ and  plot the dynamics of entanglement under different values of the quench factor $ \epsilon $. For $ \epsilon=1 $, we observe that   the Ermakov modes reduce to 
 $1$ and  consequently the entanglement (entropy) becomes constant during the dynamics.
 Now  by decreasing $ \epsilon $, we notice that the  bi-oscillations appear but for  the particular value  $ \epsilon=0.01 $   the generation of entanglement requires a time to establish.  In   the symmetric case, we conclude that the optimal value of the quench factor is $\epsilon \longrightarrow 0$ (ultra-cold  regime). However, the  optimal initial coupling  depends  on the target state $\Psi(t_{f})$ because as noticed before the coupling plays two important roles such that it can be used to engineer the time for establishing entanglement and   modulate the frequency of $ S_{v} $ oscillations. {In cooling experiments    with harmonic traps \cite{R35}, $ \epsilon $ plays a prominent role, indeed by fixing the final time $ t_{f}$ one can ask  for which value of $ \epsilon $ the entanglement will be maximal at $ t_{f} $. For instance, if we choose $ t_{f}=20$, it appears  that the optimal value of the quench is $ \epsilon= 0.1 $ and $ \epsilon=0.01 $ for the case $ t_{f}>23  $.  }
 
 \subsection{Bi-symmetric state $\sigma_{A}=\sigma_{C}\neq \sigma_{B}$ }

 We assume that our system is invariant under the permutation $A\longleftrightarrow C$,
 meaning  that $\omega_{1}(t)=\omega_{3}(t):=\omega(t)$ and $C_{12}(t)=C_{23}(t):=C(t)$, and use the notations
 $\omega_{2}(t):=\tilde{\omega}(t)$ and $ C_{13}(t):=\tilde{C}(t) $. 
 Consequently, the information content will be totally described by the both purities $P_{A}=P_{C}$ and $P_{B}$.   
 
\begin{figure}[htbphtbp]
	\centering
	\includegraphics[width=6.4cm, height=4cm]{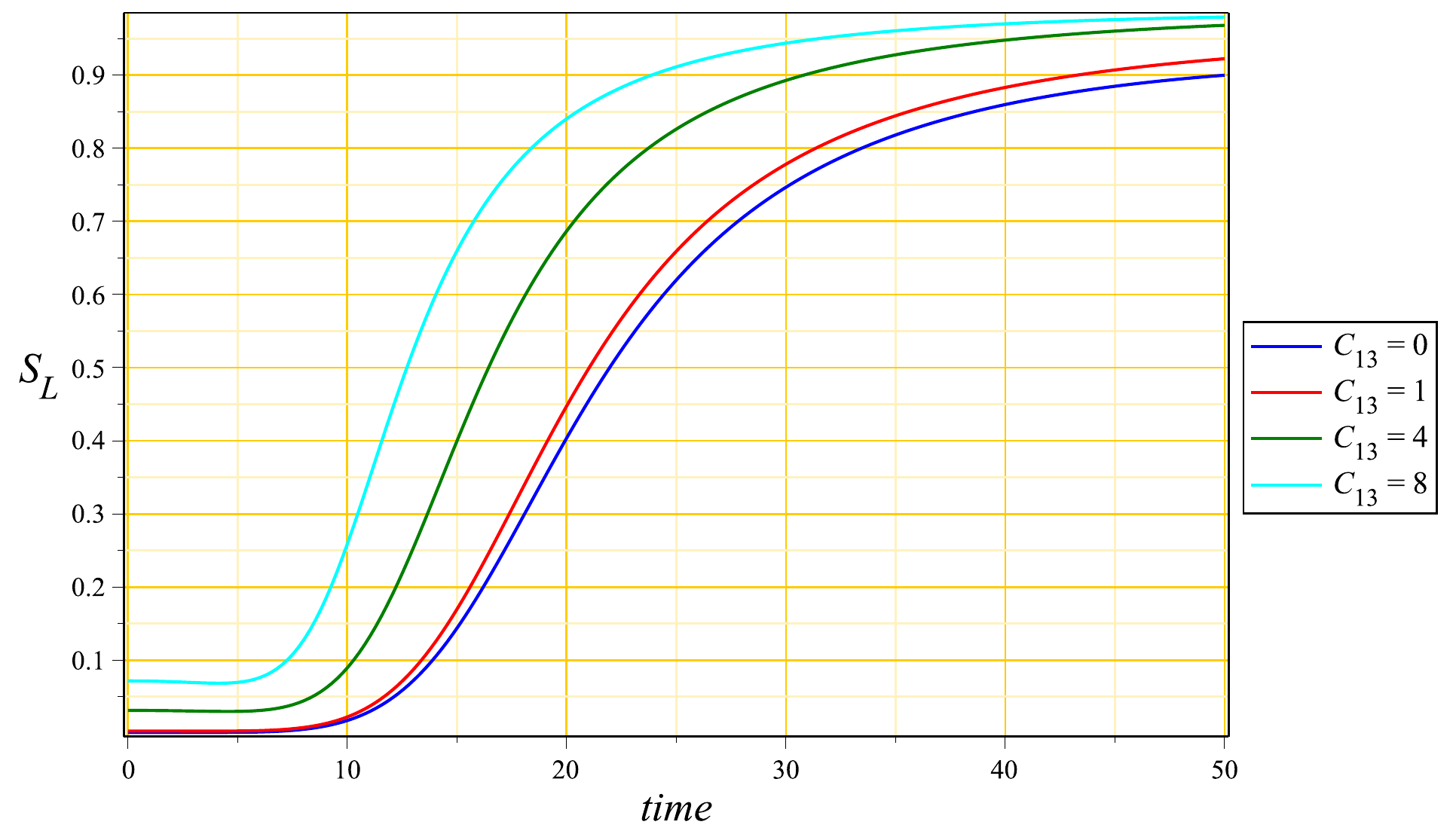}
	\hspace{2cm}
	\includegraphics[width=6.4cm, height=4cm]{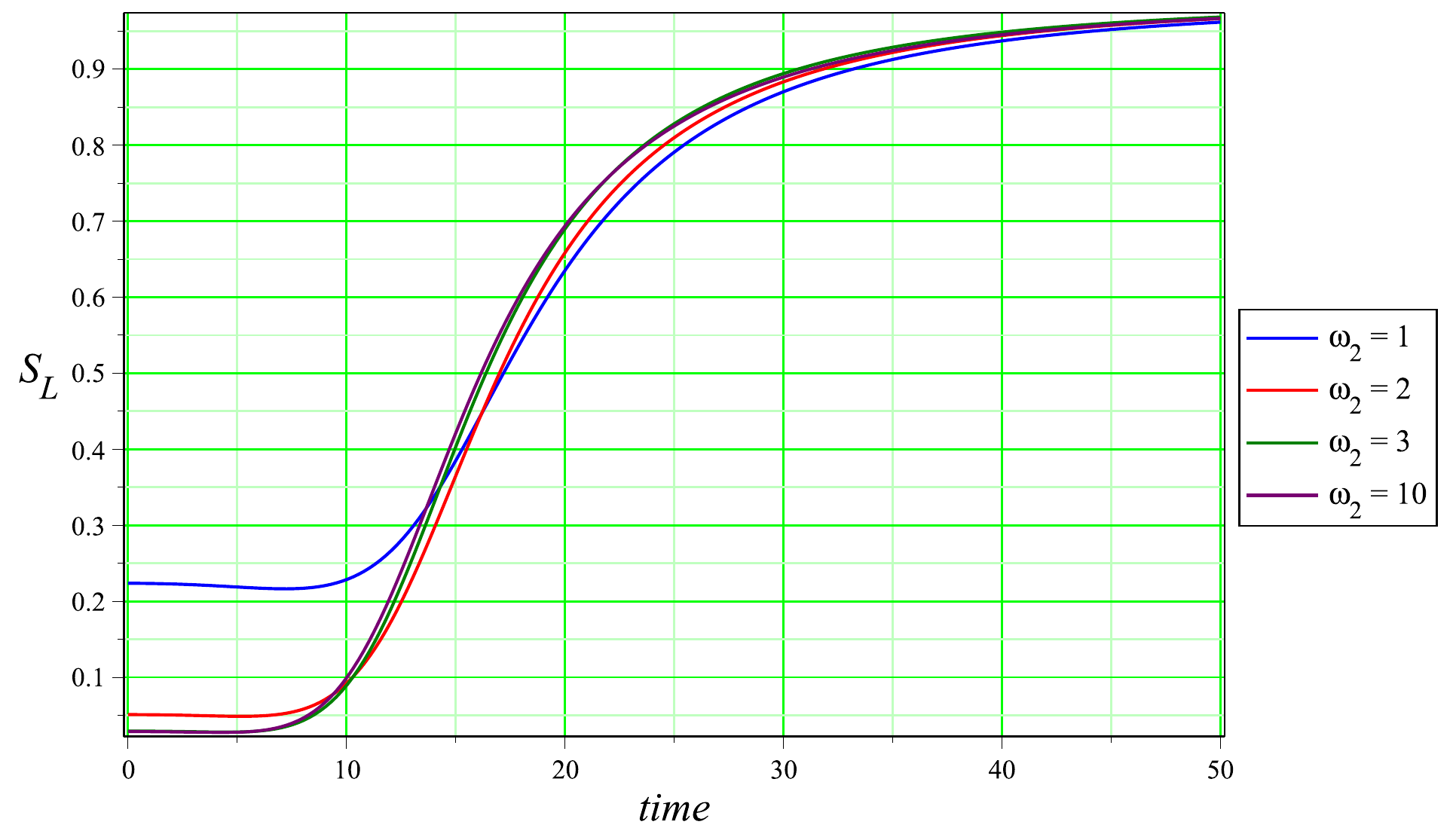}
	\captionof{figure}{\sf (color online)
		The effects of the lateral coupling $ \tilde{C}(0)$  and 
		center frequency $\tilde{\omega}(0) $  on the the dynamics of entanglement  in the time scale $[0,50]$. In  left panel: effect of $ \tilde{C}(0)$  
		for  $ C(0)=1.5$, $\omega(0)=3$,  $\tilde{\omega}(0)=5 $ and $ \epsilon=0.01$. In right panel:  effect of  $\tilde{\omega}(0) $ for $ C(0)=1.5$, $\omega(0)=5$, $ \tilde{C}(0)=4$ and $ \epsilon=0.01 $  }\label{fig3}
\end{figure}

In Figure \ref{fig3} we present the dynamics of mixedness $ S_{L}(C) $ 
versus time under suitable conditions of the lateral coupling $ \tilde{C}(0)$ and center frequency $\tilde{\omega}(0) $. Indeed, left panel  shows the effect of the mixedness for different values of $ \tilde{C}(0)$ between oscillators $ A $ and $ C $ and we choose the quench factor $ \epsilon=0.01 $. 
We observe that the amount of mixedness in monotonically increases with respect to $ \tilde{C}(0)$, which is obvious because when  $ \tilde{C}(0)$ increases the purity of the reduced state will be lost, then we have the increasing of mixedness.
On the other hand, we notice that
 the establishment of mixedness or entanglement requires a specific time, which  decreases by increasing lateral coupling. We emphasis  that the main difference between the present case and previous one is that the coupling does not affect the amplitude of mixedness but only the frequency. Here the coupling contributes in the both sides by modulating the frequency and amplitude of mixedness oscillations. 
In right panel, we show the effect of the center frequency $\tilde{\omega}(0) $ on the dynamics of $ S_{L}(C) $ under specific choice of other parameters. It is clearly seen that   $\tilde{\omega}(0) $ does not  affect the maximal value in time interval $ \left[ 0,50\right]$. Now by increasing $\tilde{\omega}(0) $, we observe that  the  amount of mixedness decreases in time interval $ \left[0,10 \right]$  and 
 $\tilde{\omega}(0) $
does not affect the time  of the entanglement establishment. 

\begin{figure}[htbphtbp]
	\centering
	\includegraphics[width=7.4cm, height=5cm]{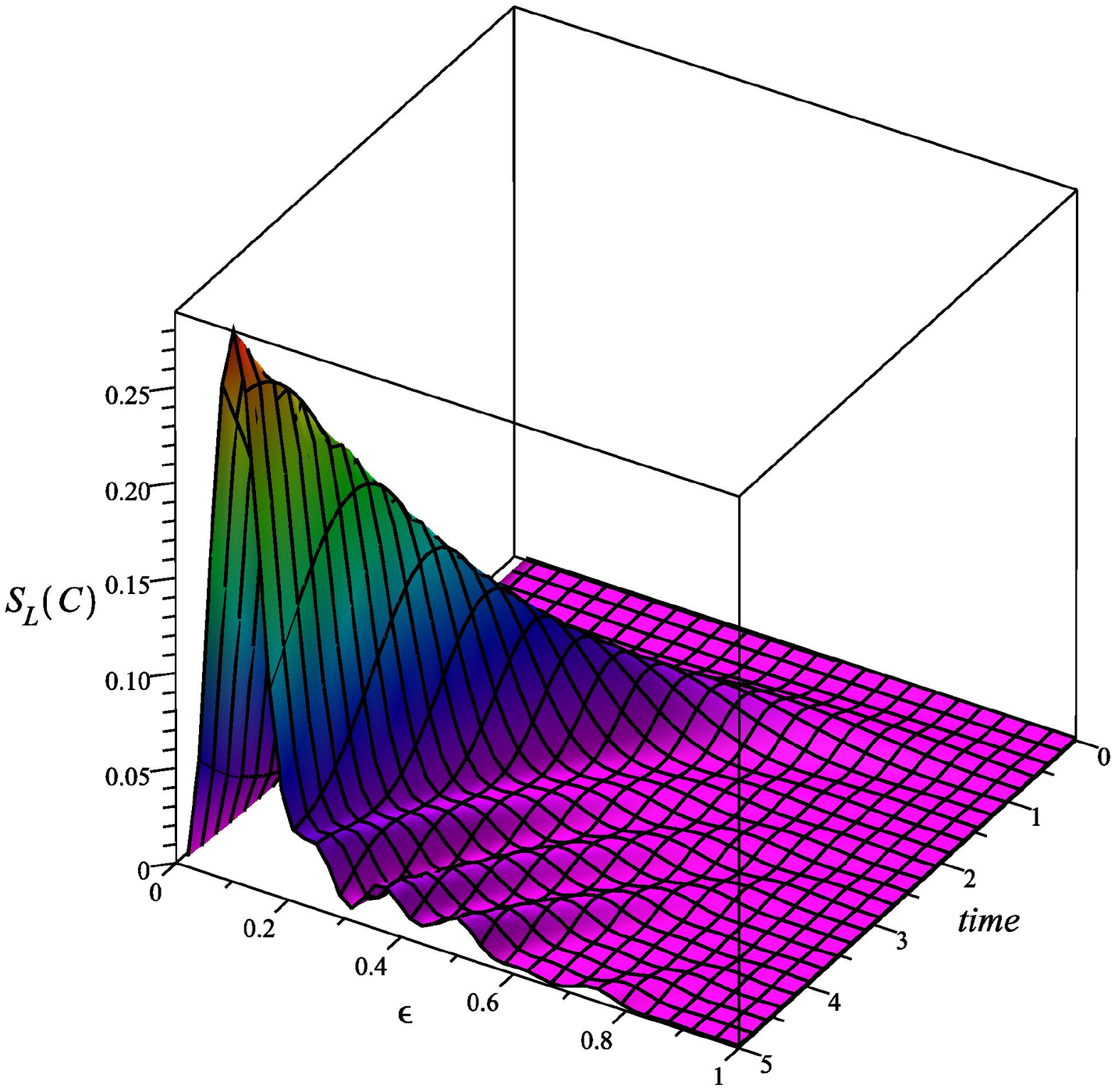}
	\includegraphics[width=7.4cm, height=5cm]{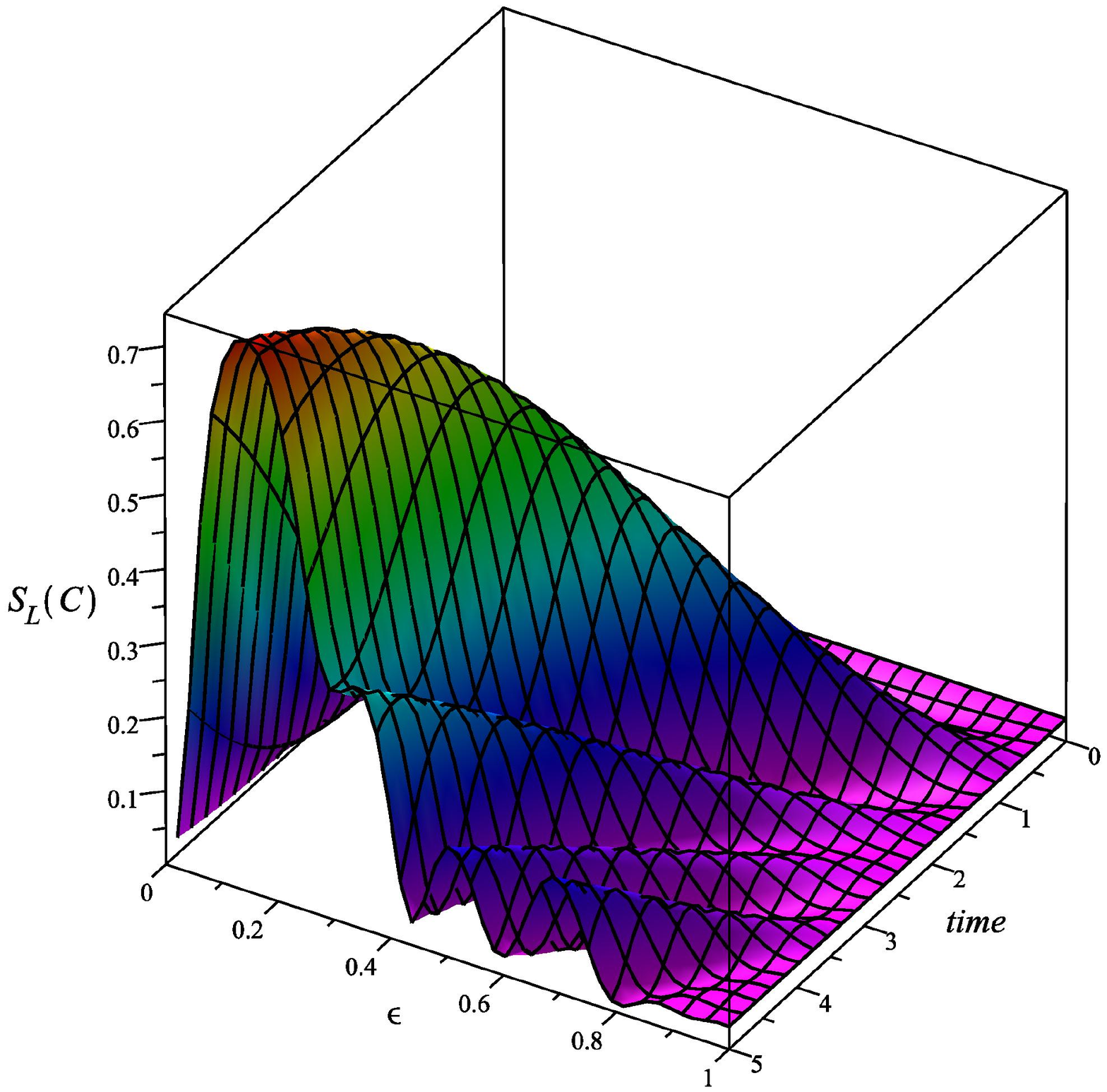}
	\captionof{figure}{\sf (color online)
		The effect of the quench factor $ \epsilon$  on the  dynamics of mixedness $ S_{L}(C) $  in the time scale $[0,5]$ for    $ C(0)=1.5$, $\omega(0)=3$,  $\tilde{\omega}(0)=5 $,  $ \tilde{C}(0)=0$ (left panel) and $\tilde{C}(0)=4$ (right  panel).   }\label{fig4}
\end{figure}

In Figure \ref{fig4}, we present a three dimension plot of the effect  of the quench factor $ \epsilon $ on the  dynamics of mixedness
by choosing the values $ C(0)=1.5$, $\omega(0)=3$ and  $\tilde{\omega}(0)=5 $ .
We observe that
the dynamics shows a critical point in the vicinity of the point $ (5, 0.25) $. It is clear that the quench factor  contributes on the modulation of the amplitude and  phase of multi-oscillations, which is a consequence of the phases $\sim  2 \epsilon \sigma_{j}(0) t$ of the three modes $\rho_{j}$ with $j=1,2,3$. Note that the distribution of entanglement between the subsystems is related to the lateral coupling $\tilde{C}(0)$, {it increases dramatically for a strong coupling i,e $\tilde{C}(0)\gg C(0)$ }.

\begin{figure}[htbphtbp]
	\centering
	\includegraphics[width=7.4cm, height=5cm]{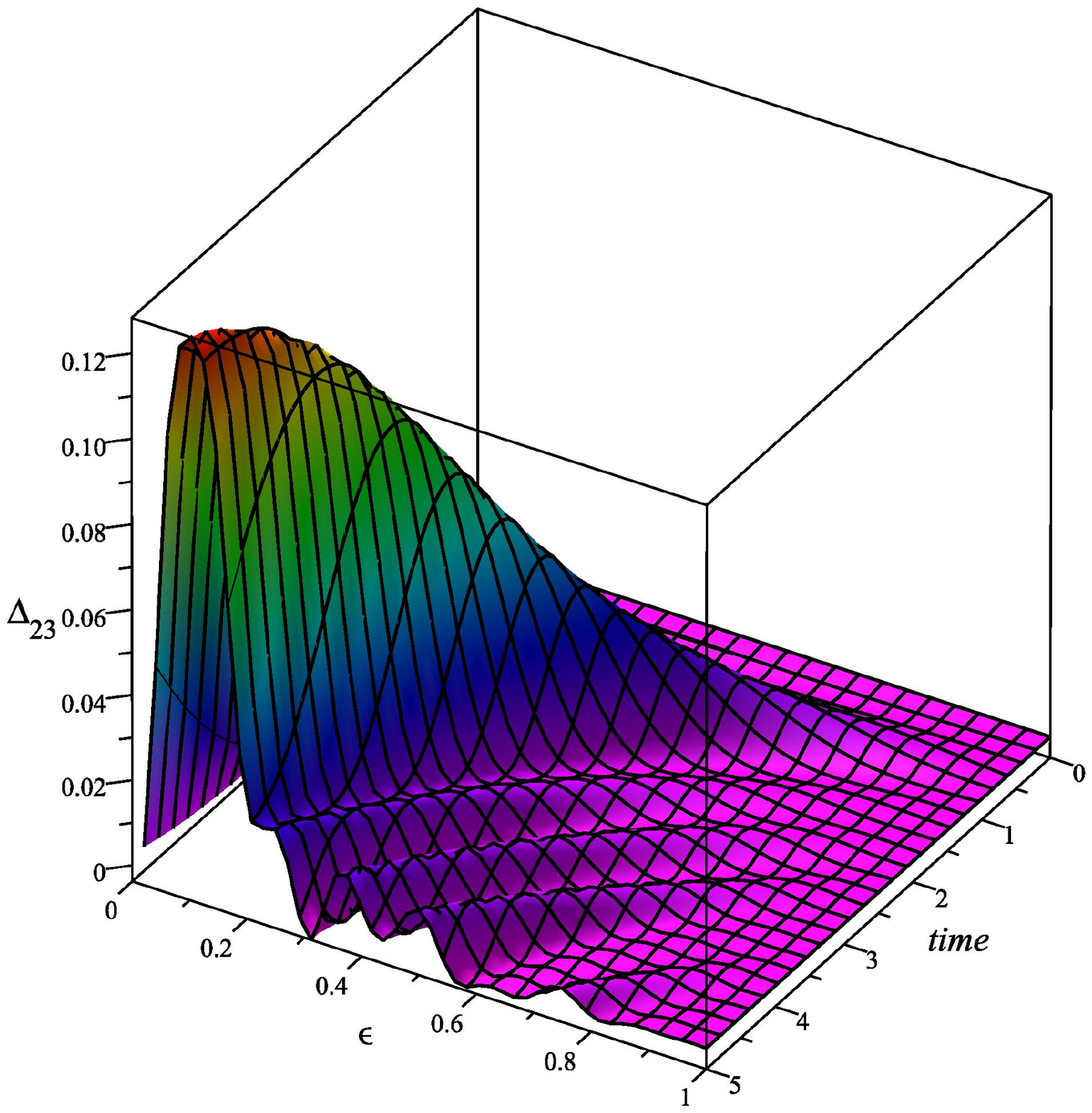}
	\includegraphics[width=5.8cm, height=5cm]{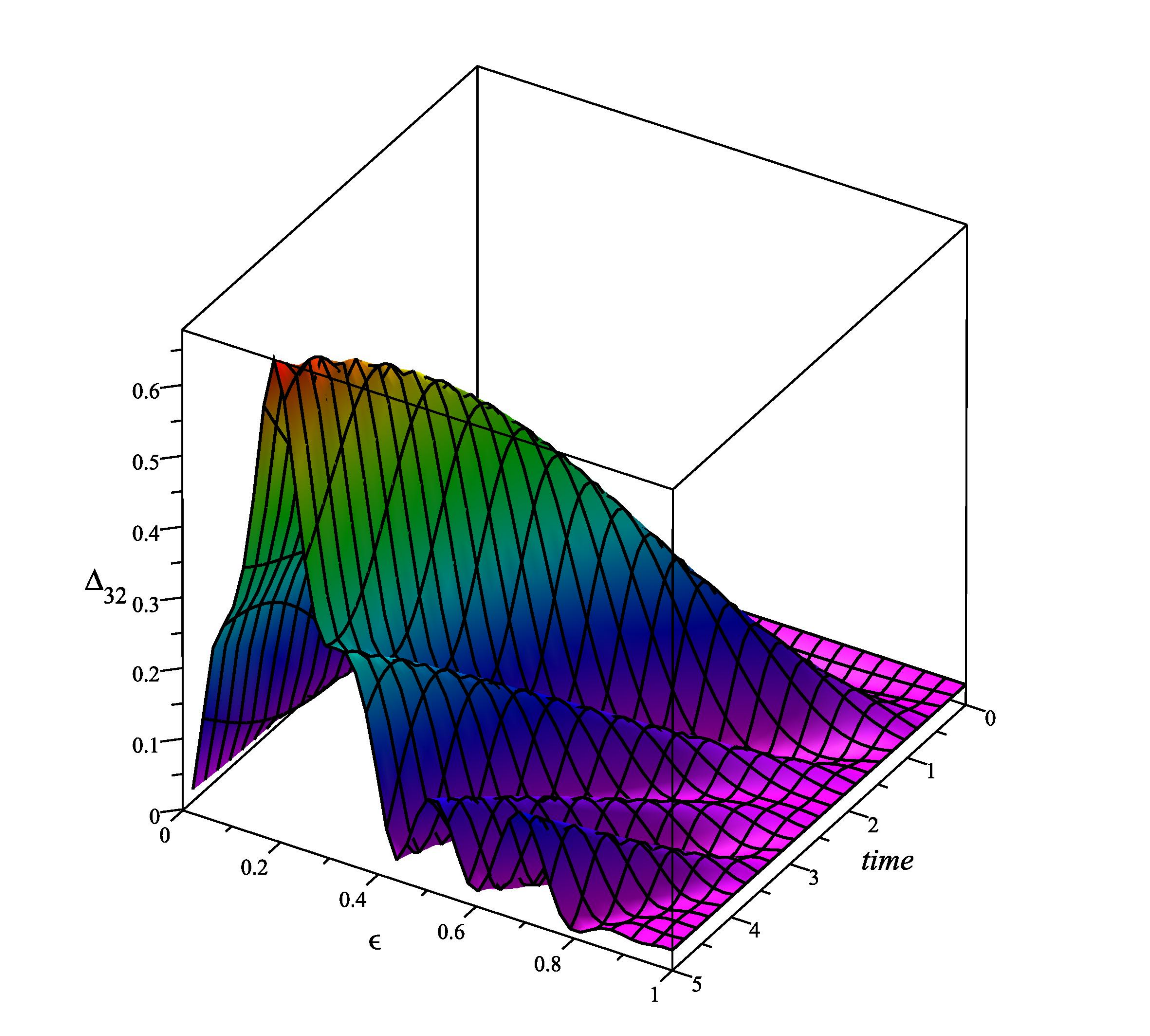}
	\captionof{figure}{\sf (color online)
		{The dynamics of the distribution of  mixedness  $\Delta_{23}:=S_{L}(B)-S_{L}(C)=- \Delta_{32}$ and the effect of the quench factor $\epsilon$ for  $C(0)=1.5 $, $\tilde{\omega}(0)=5$, $\omega(0)=3$,  $ \tilde{C}(0)=0$ (left panel) and $ \tilde{C}(0)=4$ (right panel) \label{fig6}}.}
\end{figure}

To show the effect of the quench factor $\epsilon$
on the distribution of entanglement,  we plot $ S_{L}(C)-S_{L}(B) $ versus $\epsilon$ and time    in Figure \ref{fig6}. As expected when the lateral coupling vanishes our system becomes  equivalent to an open harmonic chain \cite{R30}. Consequently, the central oscillator will encode an important amount of entanglement compared to the others. Whereas, by coupling  the lateral  oscillators $ A $ and $ C $  with the value $\tilde{C}(0)=4$,  the entanglement 
redistributes between the central and lateral oscillators. Now by increasing the quench factor $\epsilon$, the behavior approaches to the symmetric regime, namely $ S_{L}(C)-S_{L}(B)\sim 0 $ and   the critical point shifts back to  $ (t=5,\epsilon=0.2) $. 

\subsection{Fully non-symmetric state }
\begin{figure}[htbphtbp]
	\centering
	\includegraphics[width=5.5cm,height=4cm]{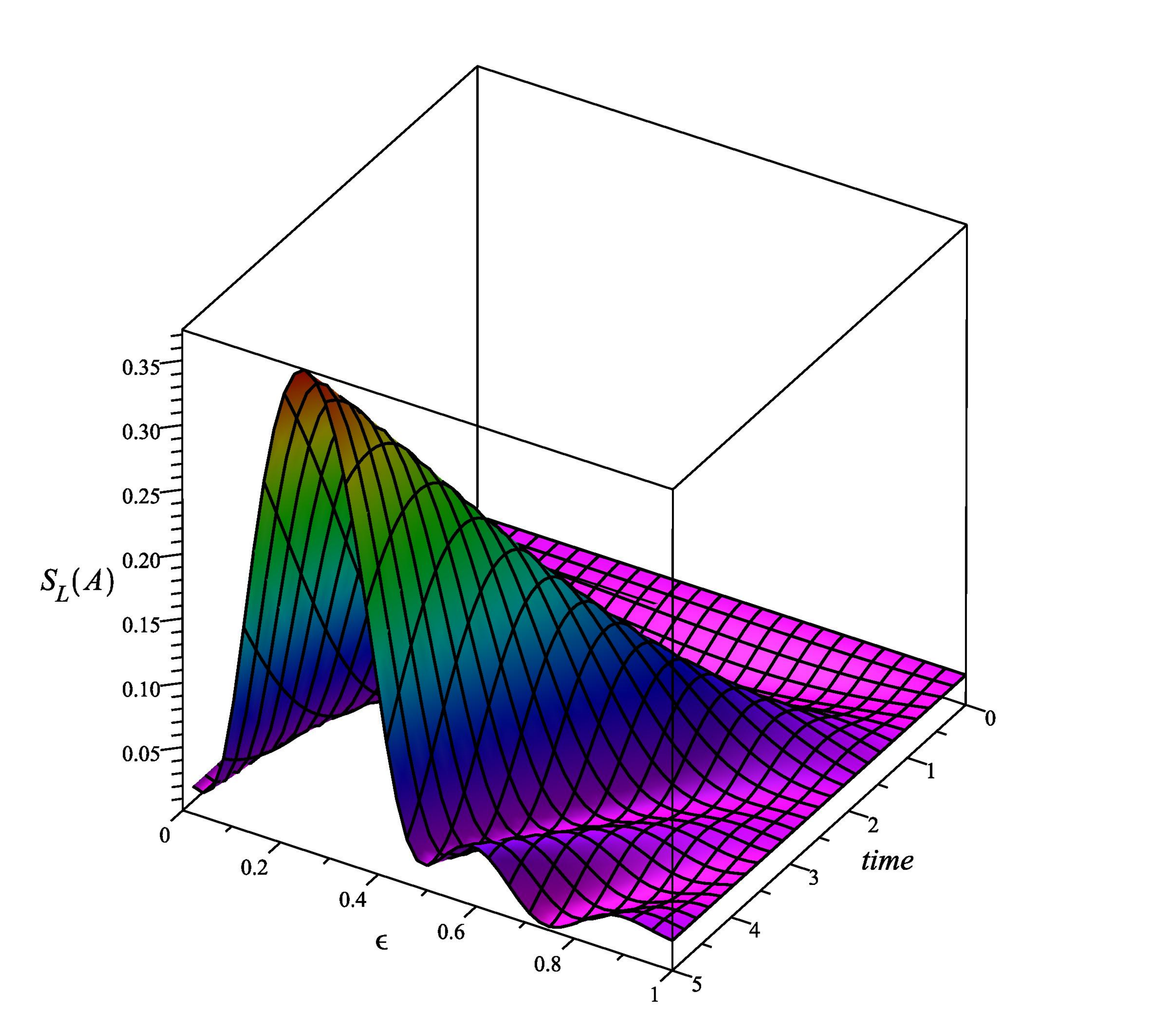}
	\includegraphics[width=5.5cm,height=4cm]{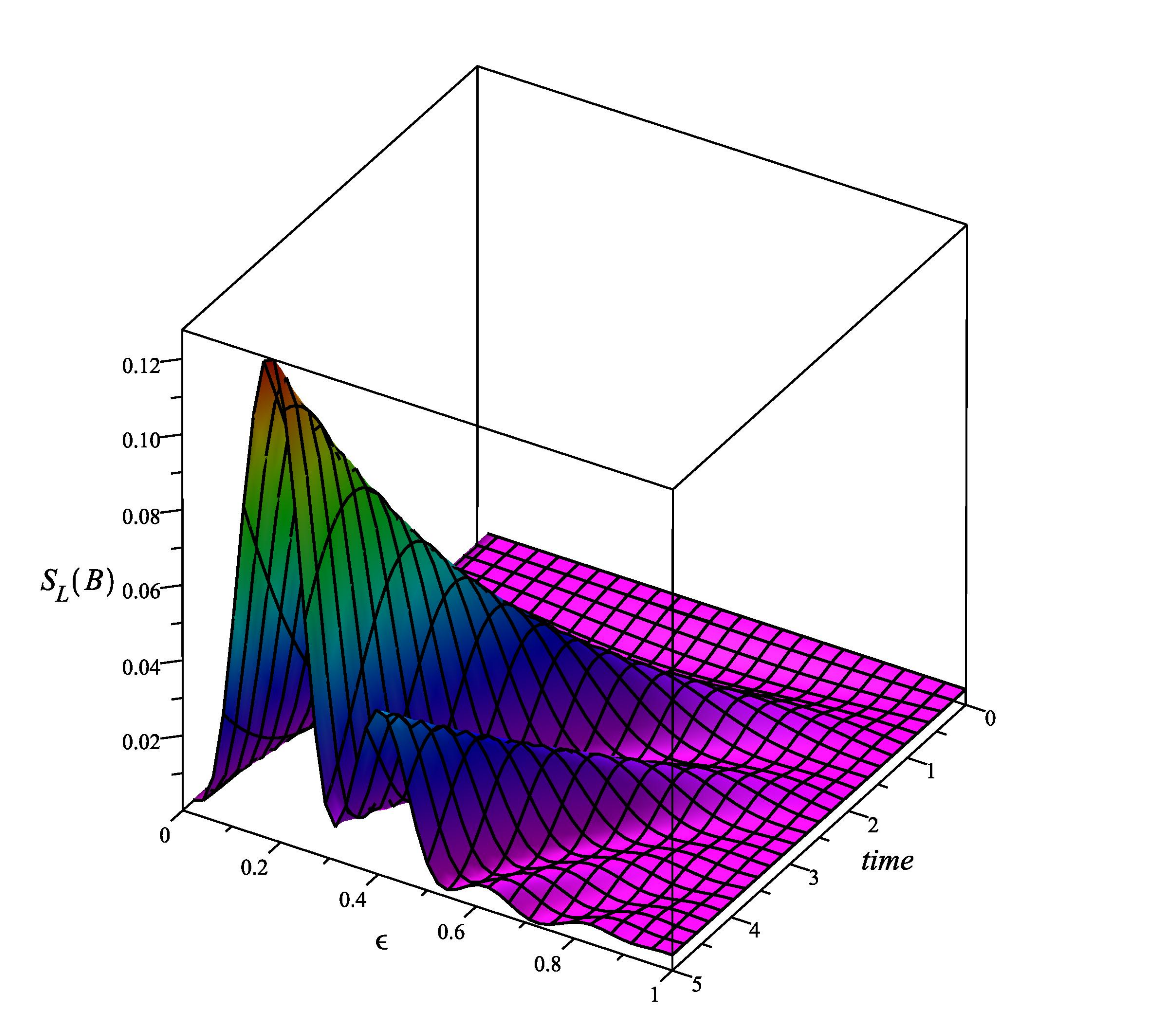}
	\includegraphics[width=5.5cm,height=4cm]{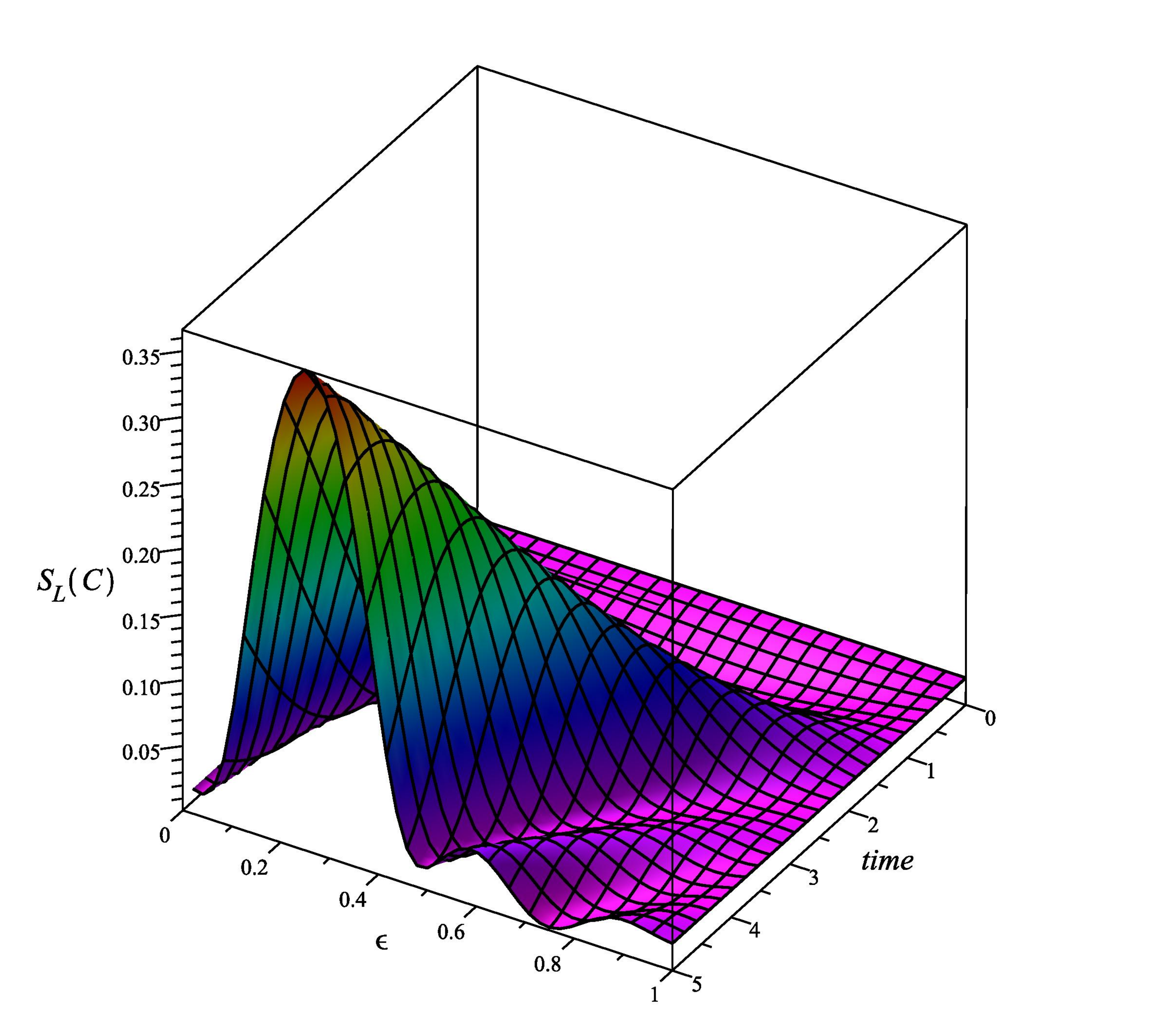}
	\captionof{figure}{\sf (color online)
	The dynamics of mixedness and the effect of the quanch factor $\epsilon $
		for two sets of the couplings   $ \left( C_{12}(0),C_{13}(0),  C_{23}(0)\right)=\left( 1,3,2\right) $ and   frequencies $\left( \omega_{1}(0),\omega_{2}(0),  \omega_{2}(0)\right) =\left( 0.5,0.8,0.35\right)$}\label{fig 88}.
\end{figure}

A state is called fully non symmetric if and only if $ \forall i<j\leq 3$ we have $ \det\sigma_{i}\neq\det\sigma_{j}$. In this case, all parameters are different
and   then the Hamiltonian is not invariant under permutation of modes. Note that the system can not be degenerate $ \forall i<j\leq 3,\,\, \sigma_{i}\neq\sigma_{j} $.
{The dynamics of mixedness of the fully non-symmetric state  is plotted in Figure \ref{fig 88}, which presents a clear hierarchy  between the central oscillator and the lateral ones $ S_{L} (A)\sim S_{L}(C)\leq S_{L}
(B)$ that strongly depends  on the quench factor $\epsilon$. It decreases as we approach
to the time-dependent regime and the optimal value  is $\epsilon\sim 0.2$.}

 \section{
 	Uncertainties, genuine  tripartite entanglement  and coherence \label{sect4}}
 \subsection{Dynamics of 
 	uncertainties and genuine  tripartite entanglement}
 
We will investigate the dynamics of partial Heisenberg uncertainties of each mode  and the genuine tripartite entanglement. For this, we transform the   covariance matrix $\sigma(t)$ to a simple form called \textit{standard form} $\tilde{\sigma}(t)$ using a  local (single mode) symplectic transformation \cite{R8, R7}. Recall  that  the quantum features (entanglement, correlations, mixedness~$\cdots$) encoded in the state under consideration are invariant. Then, the standard form $\tilde{\sigma}$ relative to our system is  given by \cite{R18}
\begin{equation}
\tilde{\sigma}(t)=
  \begin{pmatrix}
    P_{A}^{-1}(t) & 0 & C_{12}^{+}(t) & 0 & C_{13}^{+}(t) & 0 \\
    0 & P_{A}^{-1}(t)  & 0 & C_{12}^{-}(t) & 0 & C_{13}^{-}(t) \\
    C_{12}^{+}(t) & 0 & P_{B}^{-1}(t)  & 0 & C_{23}^{+}(t) & 0 \\
    0 & C_{12}^{-}(t)& 0 & P_{B}^{-1}(t)  & 0 & C_{23}^{-}(t) \\
     C_{13}^{+}(t)& 0 & C_{23}^{+}(t) & 0& P_{C}^{-1}(t)  & 0 \\
    0 & C_{13}^{-}(t) & 0 & C_{23}^{-}(t) & 0 & P_{C}^{-1}(t)  \\
  \end{pmatrix}
\end{equation}
where $P_{j}(t)$ is the local purity relative to the reduced modes $j=A,B,C$ and  $C_{ij}^{\pm}(t)
$ are correlations (classical and quantum) between  modes $i$ and $j$. The Heisenberg uncertainty associated to the reduced modes is equivalent  to 
\begin{equation}
2\Delta (x_{i})\Delta (p_{i})\geq 1\ \ \ \Leftrightarrow \ \ \ \det\sigma_{i}\geqslant 1,\qquad i=A,B,C.
\end{equation}

\begin{figure}[H]
  	\centering
  	\includegraphics[width=5.4cm,height=5cm]{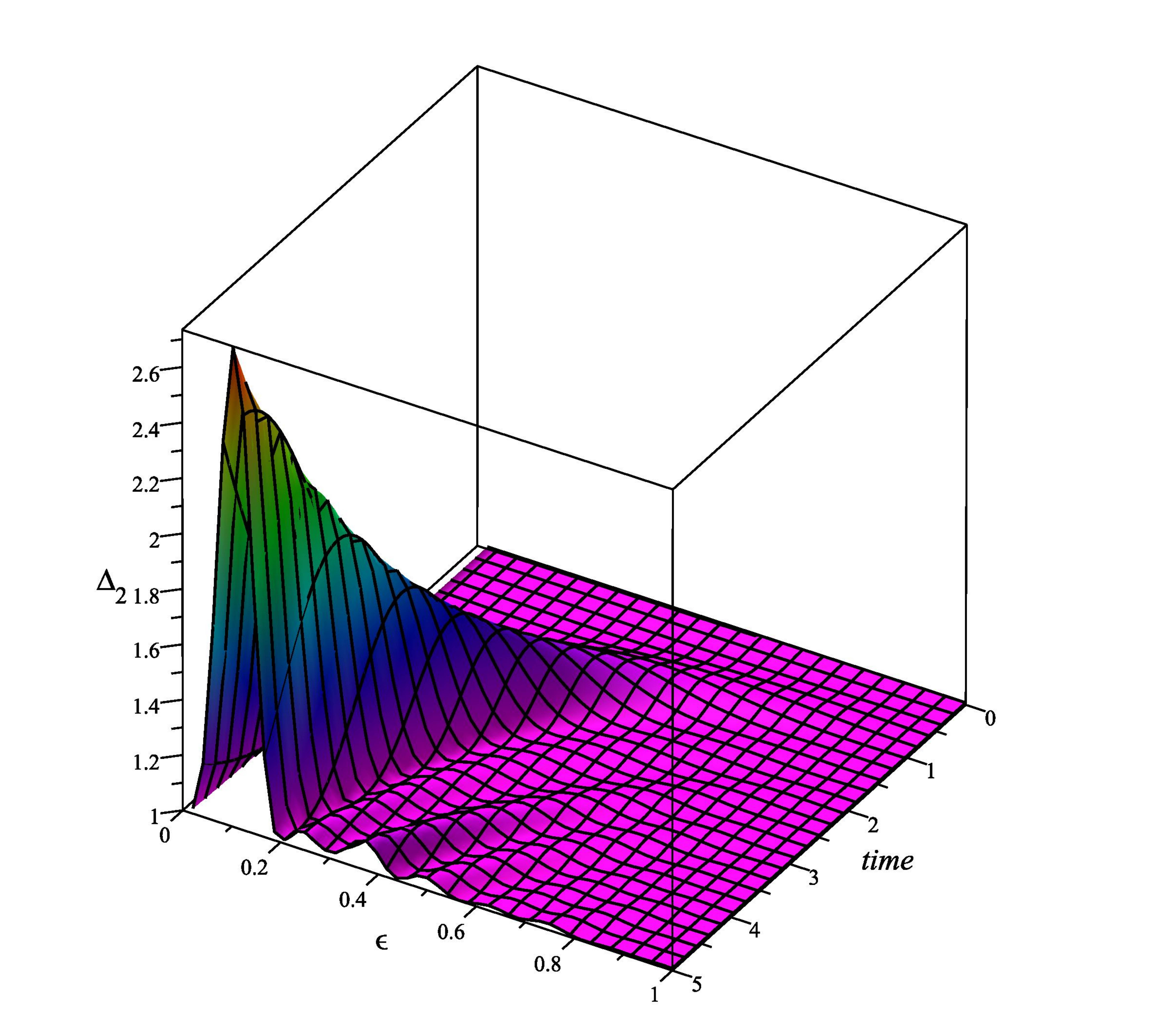}
  	\hspace{2cm}
  	\includegraphics[width=5.4cm,height=5cm]{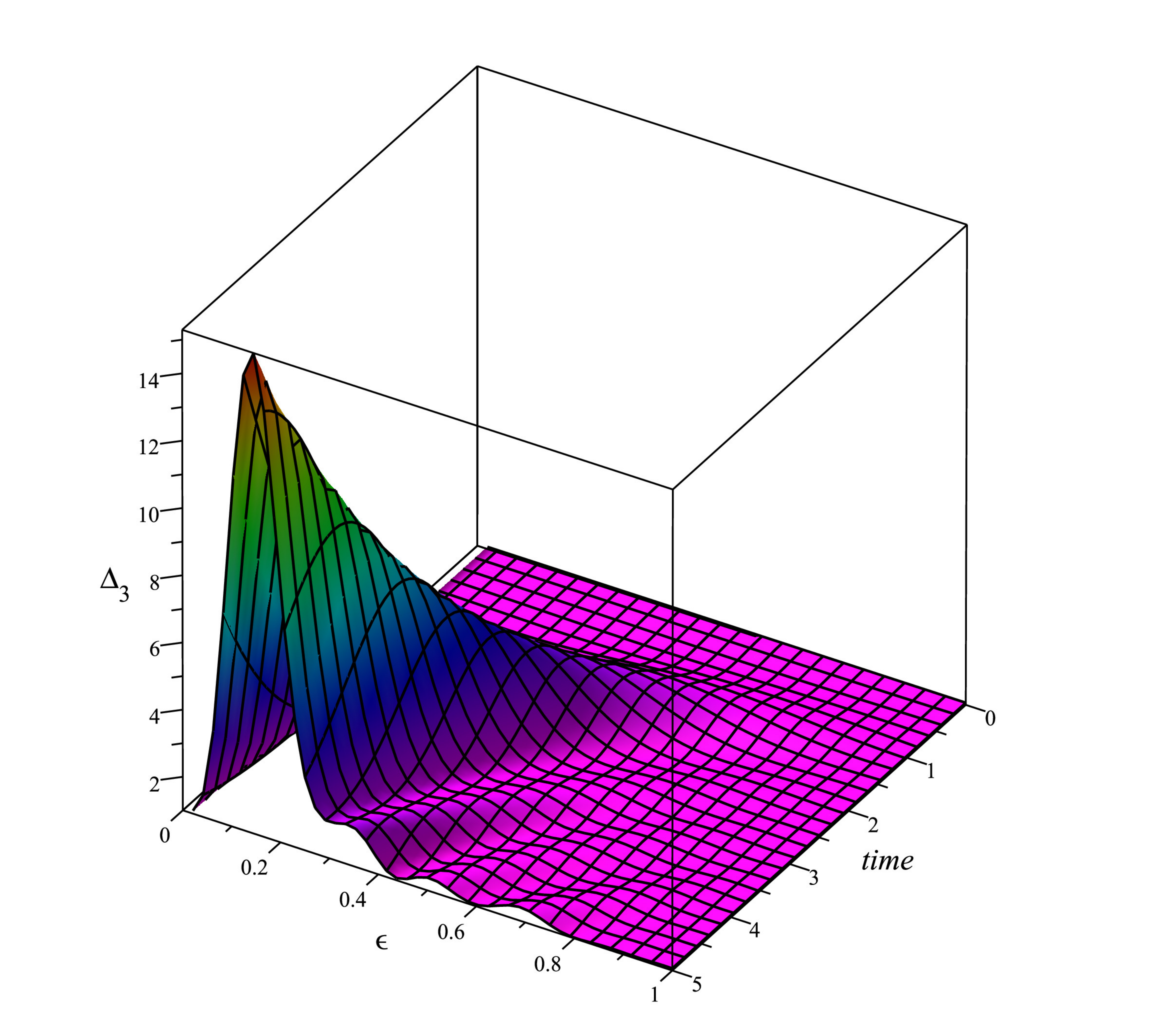}
  	\captionof{figure}{\sf (color online)
  The dynamics of the partial uncertainties $\Delta_{j}:= \Delta(x_{j})\Delta(p_{j}) $ and the effect of the quench factor $ \epsilon$ for  $C(0)=1.5 $, $\tilde{C}(0)=4$, $\tilde{\omega}(0)=5$ and $\omega(0)=3$. Dynamics of $ \Delta_{2} $ in left panel and that  of $  \Delta_{3}  $ in right panel.\label{fig5}}  
\end{figure}

 Figure \ref{fig5} presents the dynamics of uncertainties versus the quench factor $\epsilon$ and time under suitable conditions of the involved parameters.
One can notice that  the Heisenberg uncertainties show  similar behavior 
for  the three oscillators, except that the amplitude  of oscillation are different such that it is very important for the  lateral oscillators $ A $ and $ C $, which is due to the
presence of  the lateral coupling $ \tilde{C}(0) $. It is interesting to note that in the limiting case $ \epsilon\longrightarrow 1 $, the Heisenberg uncertainty  saturates. We emphasis that the uncertainty is not violated during the dynamics, which guarantees  the quantumness of the state. 
  
  To quantify the tripartite entanglement we use the Rényi entropy $ \mathcal{S}_{2} $ and  confine ourselves in the case of fully inseparable bi-symmetric state with respect to the central oscillator $ B $, and follow the dynamics of tripartite entanglement. The state in that case should verify the inseparability condition  (\ref{fs}).
  By using the bi-symmetry criterion one can show that the genuine tripartite entanglement $\mathcal{E}_{2}$ takes the form
  \begin{eqnarray}
  \mathcal{E}_{2}(t,\epsilon)=\ln\left(\dfrac{8}{P_{1}^{2}(t,\epsilon)P_{2}(t,\epsilon)h(t,\epsilon)}\right)
  \end{eqnarray}
 where we have set the quantities
  \begin{eqnarray}
  h(t,\epsilon)&=&4P_{1}^{-2}\left(1+P_{2}^{-2}\right)+P_{2}^{-2}\left(2-P_{2}^{-2}\right)-\sqrt{\delta}-1\\
  \delta(t,\epsilon)&=&\prod_{\mu,\nu=0}^{1}\left[ (1+(-1)^{\mu})P_{1}+(-1)^{\nu}P_{2})\right].
  	\end{eqnarray}

  \begin{figure}[H]
  	\centering
  	\includegraphics[width=5.4cm,height=3.5cm]{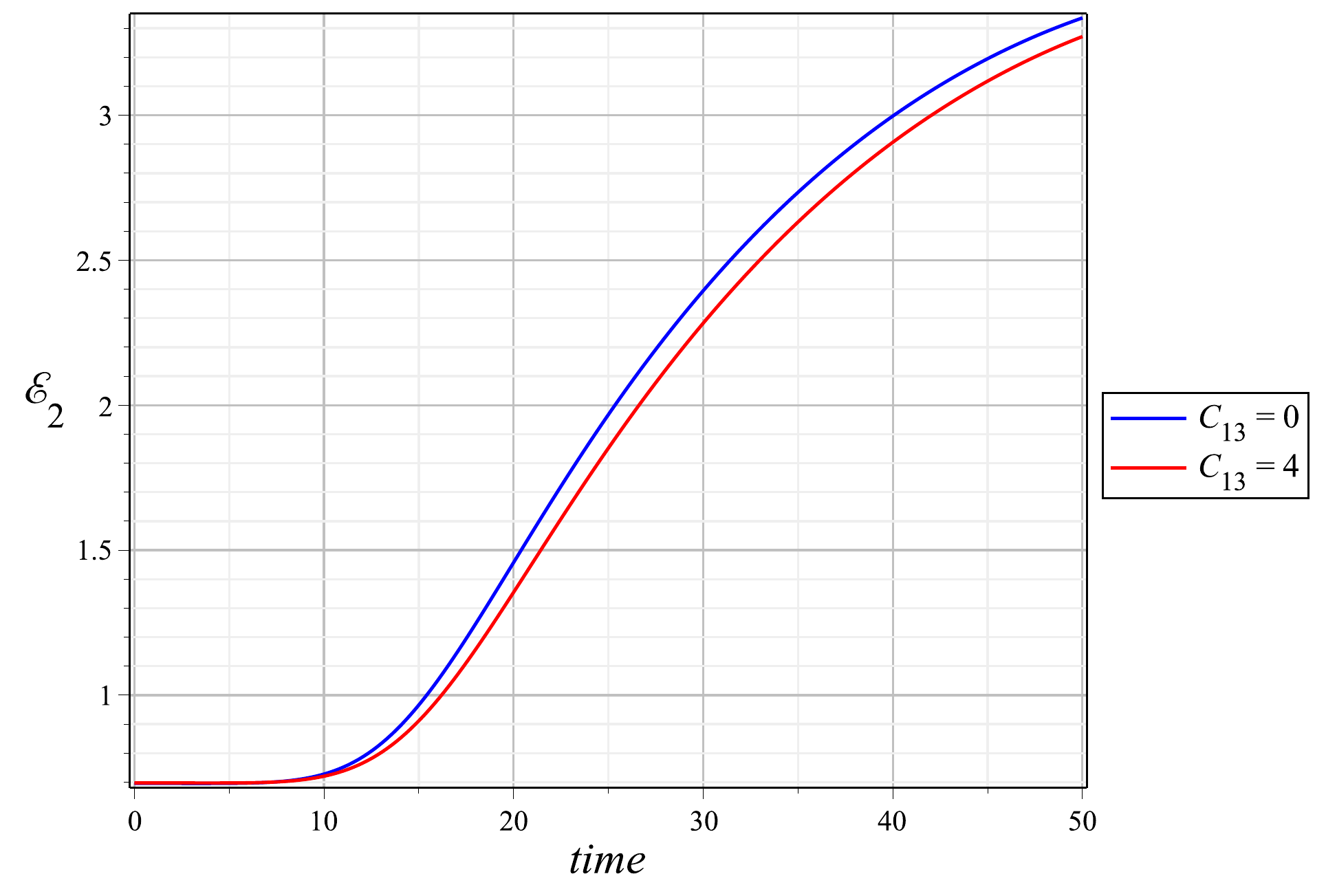}
  	\hspace{2cm}
  	\includegraphics[width=5.4cm,height=3.5cm]{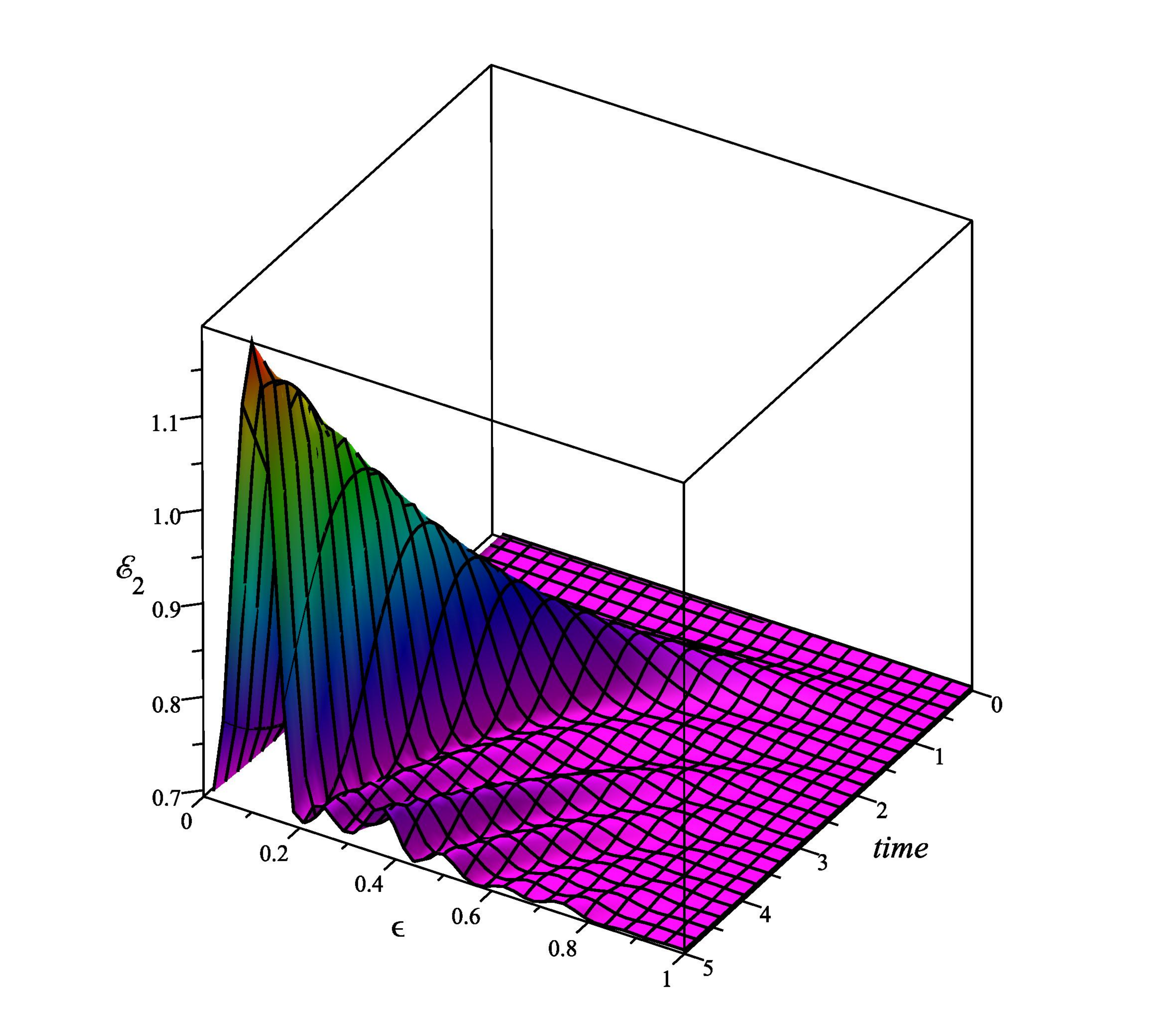}
  	\captionof{figure}{\sf (color online)
  		The dynamics of tripartite entanglement $ \mathcal{E}_{2} $ 
  		versus time for two values of the lateral coupling $ C_{13}$ 
  		(left panel with $ \epsilon=0.01 $).
  		The effect of quench parameter $\epsilon$ on the dynamics of $ \mathcal{E}_{2} $
  		(right panel with $ C_{13}=4 $).} \label{tripa}
  \end{figure}
  	In {Figure} \ref{tripa}, we plot  the dynamics of  genuine tripartite entanglement $ \mathcal{E}_{2} $ versus time for some values of the involved parameters. In left panel, we remark that the generation of $ \mathcal{E}_{2} $ requires a specific time to be established. By increasing the  lateral coupling $ C_{13} $, $ \mathcal{E}_{2} $  decreases in the time scale $ [0,50] $. On the other hand, when the Ermakov modes increase the coupling modulates the frequency and  amplitude of the oscillations. In right panel, we plot the dynamics in the  time scale $ [0,5] $ in order to easily  investigate the effect of the  quench factor $\epsilon$. We observe that   the optimal behavior of $ \mathcal{E}_{2} $ 
  	is obtained in the limiting case $\epsilon\longrightarrow 0.1$.  It is clearly seen that by approaching to the time-independent Hamiltonian regime $ \mathcal{E}_{2}$ becomes  constant. Note that, we have a similar dynamics regarding  the uncertainties and mixdness. 

\subsection{Dynamics of coherence}

Coherence is the principal ingredient to observe interference and is the  quantum feature key to explain several phenomena ranging from quantum optics to quantum information and quantum biology \cite{R38}. Our aim here is to show the effect of the dynamics on the generation of coherence. Since our state $\rho$ is Gaussian with  zero first  moment and  a second moment $\sigma(t)$ (\ref{si}), then $\rho$ is said to be incoherent if it is diagonal when expressed in a fixed orthonormal basis. A suitable measure of coherence $C(\rho)$ must verify the following postulates \cite{R24,R25}: 
\begin{itemize}
\item $P_{1}$: $C(\rho)\geq 0$ and $C(\rho)=0$ if and only if $\rho\in \mathcal{I}$, with $\mathcal{I}$:= the set of incoherent states.
\item $P_{2}$: Non increasing under a mixture of quantum states: $\sum\limits_{n} p_{n}C(\rho_{n})\geq C\left( \sum\limits_{n}p_{n}\rho_{n}\right)$: convexity.
\item $P_{3}$: Monotonicity under  incoherent quantum operations completely positive trace preserving (ICPTP) operations:
\begin{equation}
\phi_{ICPTP}:\ \  \rho\rightarrow \sum\limits_{n} K_{n} \rho K_{n}^{+}, \qquad \sum\limits_{n}K_{n}^{+}K_{n}=\mathbb{I}
\end{equation}
$K_{n}$ are Kraus operators that stabilize the set $\mathcal{I}$ ($\forall n$ $K_{n}\mathcal{I}K_{n}\subset \mathcal{I}$ ): $C(\rho)\geq C(\phi _{ICPTP}(\rho))$.  
\end{itemize}
  Consequently, the suitable coherence measure is obtained by minimizing the geometric distance between the state 
  $\rho$ from the set of incoherent states $\mathcal{I}$, which is just for the set of  locally thermal state (tensor product of thermal state) \cite{R26}.  The global coherence is quantified as 
  \begin{eqnarray}
  C(\rho)&:=&\min\limits_{\delta\in \mathcal{I}} S\left(\rho||\delta\right)= S(\rho||\rho_{diag})= S(\rho_{diag})-S(\rho)\\ \nonumber  &\ =& -S(\rho)+\sum\limits_{i=1}^{3}\left[(\overline{n}_{i}+1)\ln(\overline{n}_{i}+1)-\overline{n}_{i}\ln(\overline{n}_{i})\right] \nonumber
 \end{eqnarray}
where $ \overline{n}_{i} $ is  the mean population relative to mode $i$ and $ S(\rho) $ stands for the global von Neumann entropy which is zero because the state is pure. It follows that the three symplectic eigenvalues are equal to unity, i.e.  $\forall i,\,\,\nu_{i}=1$. Then,
 we have 
 \begin{equation}
 S(\rho)=-\sum\limits_{i=1}^{3}\left[\frac{\nu_{i}-1}{2}\ln\frac{\nu_{i}-1}{2}-\frac{\nu_{i}+1}{2}\ln\frac{\nu_{i}+1}{2}\right]=0
 \end{equation} 
 where $i=1,2,3$ denotes the  mode numbers $(A,B,C)$. To compute the coherence resource encoded in our system we begin by computing the covariance matrix in the Fock basis. This is very important because the coherence is basis dependent at variance with entanglement \cite{R38}. More precisely, the quantification of coherence requires a change of basis from the quadrature $ Q=(x_{1},p_{1},x_{2},p_{2},x_{3},p_{3}) $ to the vector $\mathbb{Q}=(A_{1},A_{1}^{+},A_{2},A_{2}^{+},A_{3},A_{3}^{+} )$, where $ q_{j}=A_{j}+A_{j}^{+} $ and $ p_{j}=-i(A_{i}-A_{i}^{+}) $. We show that the linear operator $ \mathbb{K} $ behind the change is
 \begin{eqnarray}
 \mathbb{K}=\bigoplus\limits_{j=1}^{3}\Xi, \qquad \Xi= \begin{pmatrix}
 1 & 1 \\ 
 -i & i
 \end{pmatrix} 
 \end{eqnarray}
 and therefore   the covariance matrix $ \sigma $ transforms  in the new basis according to the relation
 \begin{equation}
  \sigma'=\mathbb{K} \sigma \mathbb{K}^{+}.
 \end{equation}
 In addition, we show that the mean population $  \overline{n}_{i}$  reads as
 \begin{eqnarray}
  \overline{n}_{j}=\frac{1}{2}(\mathbb{G}_{2j,2j}+\mathbb{G}_{2j-1,2j-1}-1),\qquad j=1,2,3.
 \end{eqnarray}
 \begin{figure}[htbphtbp]
 	\centering
 	\includegraphics[width=6.4cm,height=5cm]{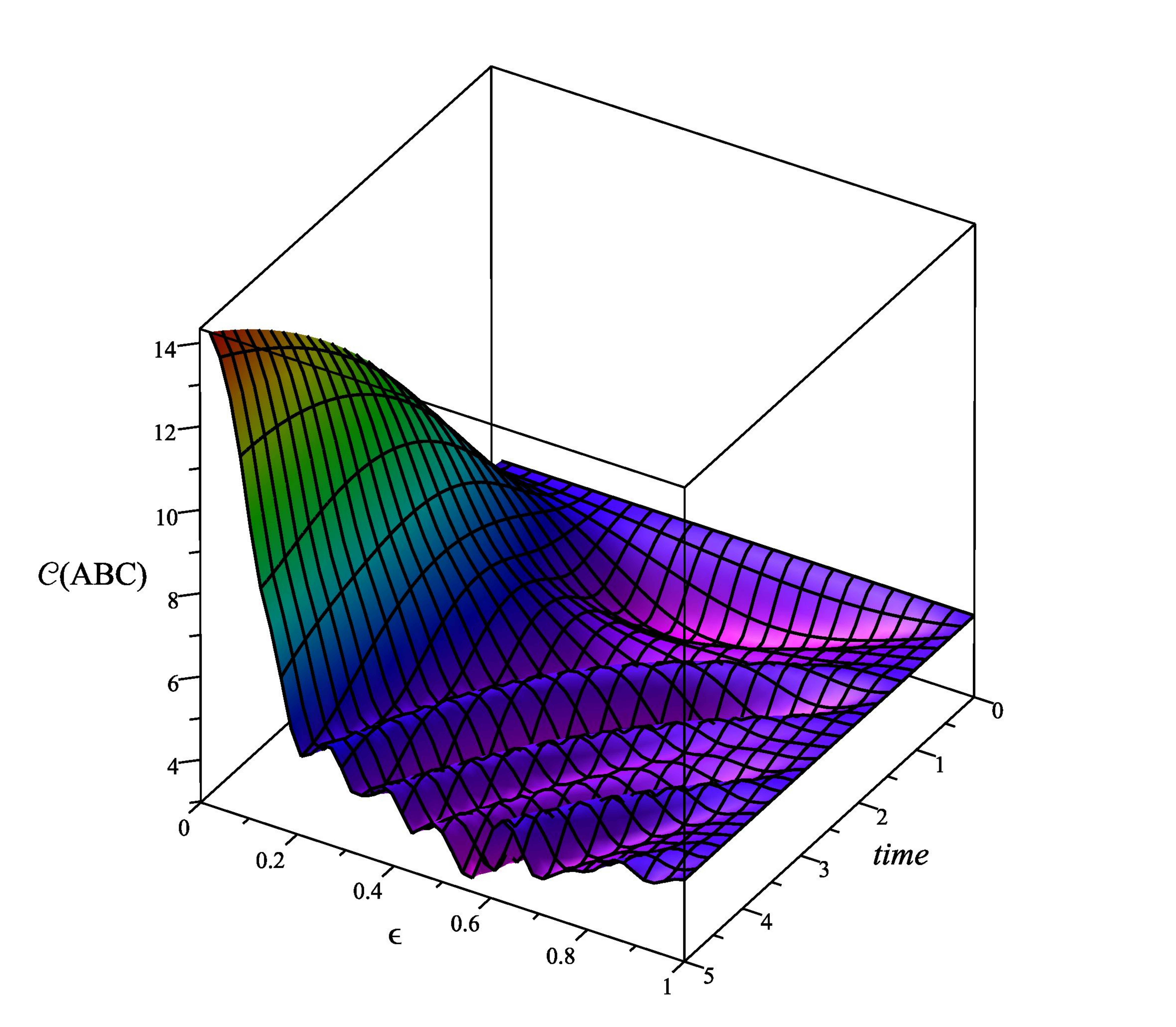}
 	\captionof{figure}{\sf (color online)
 		The dynamics of the global coherence and the effect of the quench factor $\epsilon$ for    $\tilde{\omega}(0)=5$, $\tilde{C}(0)=0$, $C(0)=1.5 $ and $ \omega(0)=3 $.\label{fig7} }
 \end{figure}

 		For the dynamics of global coherence, we plot the global coherence $ C(\rho) $ versus time scale and the quench factor $\epsilon$ in Figure \ref{fig7} for the lateral coupling $\tilde{C}(0)=0$. At variance with entanglement dynamics coherence amount does not require a delay time to be established and   the state is initially  coherent, i.e. $ C(\rho)\neq 0 $. The coherence amount exceeds the tripartie entanglement $ \mathcal{E}_{2}$ during the dynamics, which is consistent with the coherence theory \cite{R38}. The dynamics shows that the optimal coherence is obtained in the vicinity of the point $ (\epsilon=0,t=5) $. This entails that the dynamics is important to engineer the optimality of coherence encoded in the state. By increasing the quench factor to unity (time-independent regime), we observe that the coherence amount becomes constant, which due to the reduction of the  Ermakov modes to unity.

\section{Homodyne detection of highest frequency mode and redistribution of resources \label{sect5}}
\hspace{3mm}Our task  is to proof how to redistribute the resources of entanglement and coherence encoded in the reduced mode. We achieve our goal by performing a perfect homodyne detection (efficiency $\eta=1$) on the central mode $ B $. For  simplicity,  we assume that our state is bi-symmetric with respect to the detected mode and structure the state as  
\begin{eqnarray}
\sigma(t)=
  \begin{pmatrix}
    \mathcal{A}(t) & \mathcal{C}(t) \\
    \mathcal{C}^{T}(t)  & \mathcal{B}(t)  \\
  \end{pmatrix}
\end{eqnarray}
where the matrices $\mathcal{A}(t)$ and $\mathcal{B}(t)$ are the reduced $AC$ and $B$ states, respectively, while the submatrix $\mathcal{C}(t)$ contains all the correlations among the $AC$ and $B$ subsystems. Recall that, by performing a perfect  homodyne detection  of the quadrature $x_{2}$, the output state of the two mode state $AC$ will be \cite{R25,R40}
\begin{equation}
\sigma_{AC}^{out|x_{2}}:=\mathcal{A}(t)-\mathcal{C}(t)(\pi_{x_{1}}\mathcal{B}(t)\pi_{x_{2}})^{-1}\mathcal{C}^{T}(t)
\end{equation}
such that the projector  $\pi_{x_{2}}=
  \begin{pmatrix}
    1 & 0 \\
    0 & 0  \\
  \end{pmatrix}$
   is singular ($\det\pi_{x_{2}}=0$), which makes the matrix $\pi_{x_{2}}\mathcal{B}(t)\pi_{x_{2}}$ also singular. The inverse does not exist
   and therefore we use the pseudo inverse of Moore-Penrose to compute the resulting state $\sigma_{C}^{out|x_{2}}$. After a straightforward algebra, we get the following explicit expression  
\begin{equation}
\sigma_{AC}^{out|x_{2}}= \begin{pmatrix} \mathbb{G}_{22}-{\frac {{\mathbb{G}_{{24}}}^{2}}{\mathbb{G}_{{44}}
}}&-\mathbb{G}_{{12}}+{\frac {\mathbb{G}_{{24}}\mathbb{G}_{{23}}}{\mathbb{G}_{{44}}}}&\mathbb{G}_{{26}}-{\frac {\mathbb{G}_{
{24}}\mathbb{G}_{{46}}}{\mathbb{G}_{{44}}}}&-\mathbb{G}_{{25}}+{\frac {\mathbb{G}_{{24}}\mathbb{G}_{{36}}}{\mathbb{G}_{{44}}
}}\\ \noalign{\medskip}-\mathbb{G}_{{12}}+{\frac {\mathbb{G}_{{24}}\mathbb{G}_{{23}}}{\mathbb{G}_{{44}}}}&
\mathbb{G}_{{11}}-{\frac {{\mathbb{G}_{{23}}}^{2}}{\mathbb{G}_{{44}}}}&-\mathbb{G}_{{25}}+{\frac {\mathbb{G}_{{24}}\mathbb{G}_{{36}}}{\mathbb{G}_{{44}}
}}&\mathbb{G}_{{15}}-{\frac {\mathbb{G}_{{23}}\mathbb{G}_{{36}}}{\mathbb{G}_{{44}}}}
\\ \noalign{\medskip}\mathbb{G}_{{26}}-{\frac {\mathbb{G}_{
{24}}\mathbb{G}_{{46}}}{\mathbb{G}_{{44}}}}&-\mathbb{G}_{{25}}+{\frac {\mathbb{G}_{{24}}\mathbb{G}_{{36}}}{\mathbb{G}_{{44}}
}}&\mathbb{G}_{{66}}-{\frac {{\mathbb{G}_{{46}}
}^{2}}{\mathbb{G}_{{44}}}}&-\mathbb{G}_{{56}}+{\frac {\mathbb{G}_{{46}}
\mathbb{G}_{{36}}}{\mathbb{G}_{{44}}}}
\\ \noalign{\medskip}-\mathbb{G}_{{25}}+{\frac {\mathbb{G}_{{24}}\mathbb{G}_{{36}}}{\mathbb{G}_{{44}}
}} &\mathbb{G}_{{15}}-{\frac {\mathbb{G}_{{23}}\mathbb{G}_{{36}}}{\mathbb{G}_{{44}}}}&-\mathbb{G}_{{56}}+{\frac {\mathbb{G}_{{46}}
\mathbb{G}_{{36}}}{\mathbb{G}_{{44}}}}&\mathbb{G}_{{55}}-{\frac {{\mathbb{G}_{{36}}}^{2}}{\mathbb{G}_{{44}}}}
\end {pmatrix}. 
\end{equation}
Then the two new single mode purities become 
\begin{equation}
(P_{A}^{out}(t,\epsilon))^{-2}=P_{A}^{-2}(t,\epsilon)+R_{A}(t,\epsilon),\qquad (P_{C}^{out}(t,\epsilon))^{-2}=P_{C}^{-2}(t,\epsilon)+R_{C}(t,\epsilon)
\end{equation}
where the time-dependent  shifts $R_{A,C}(t,\epsilon)$ are 
\begin{eqnarray}
&&
R_{A}(t,\epsilon)=-\frac{\mathbb{G}_{11}\mathbb{G}_{24}^{2}+\mathbb{G}_{22}\mathbb{G}_{23}^{2}-2\mathbb{G}_{12}\mathbb{G}_{24}\mathbb{G}_{23}}{\mathbb{G}_{44}}\\
&& R_{C}(t,\epsilon)=-\frac{\mathbb{G}_{55}\mathbb{G}_{46}^{2}+\mathbb{G}_{66}\mathbb{G}_{36}^{2}-2\mathbb{G}_{56}\mathbb{G}_{46}\mathbb{G}_{36}}{\mathbb{G}_{44}}.
\end{eqnarray} 
Since the modes $ A $ and $ C $ are symmetric  then we have $ \Upsilon_{AB}=\Upsilon_{BC}  $ and therefore
the present  homodyne measurement  does not affect the symmetry of the state, i.e.
$ R_{A}=R_{C}$. 
It is important to note that the computation of  coherence encoded in the output state  requires  the expression of covariant matrix in the basis  $ \left\lbrace A,A^{+}\right\rbrace  $. Consequently,  the entanglement and coherence  of the reduced modes  $A$ or $C$ after measurement are,  respectively,  given by 
\begin{eqnarray}
&& S_{v}^{m|out}=\frac{\mathfrak{a}^{m|out}+1}{2}\ln\left(\frac{\mathfrak{a}^{m|out}+1}{2}\right)-\frac{\mathfrak{a}^{m|out}-1}{2}\ln\left(\frac{\mathfrak{a}^{m|out}-1}{2}\right)\\
&& C(\rho^{m|out})=-S(\rho^{m|out})+(\overline{n}_{m}^{out}+1)\ln(\overline{n}_{m}^{out}+1)-\overline{n}_{m}^{out}\ln(\overline{n}_{m}^{out})
\end{eqnarray}
and we have set the quantities
\begin{eqnarray}
\overline{n}_{m}^{out}=\frac{1}{2}\left(\mathbb{G}_{22}+\mathbb{G}_{11}-{\frac {{\mathbb{G}_{{24}}}^{2}+{\mathbb{G}_{{23}}}^{2}}{\mathbb{G}_{{44}}
}} -1\right), \qquad  \mathfrak{a}^{m|out}=\frac{1}{P^{m|out}}={\sqrt{det(\sigma_{m}^{out})}}
\end{eqnarray}
where $ m=A,C.$

\begin{figure}[H]
	\centering 
	\includegraphics[width=6.5cm]{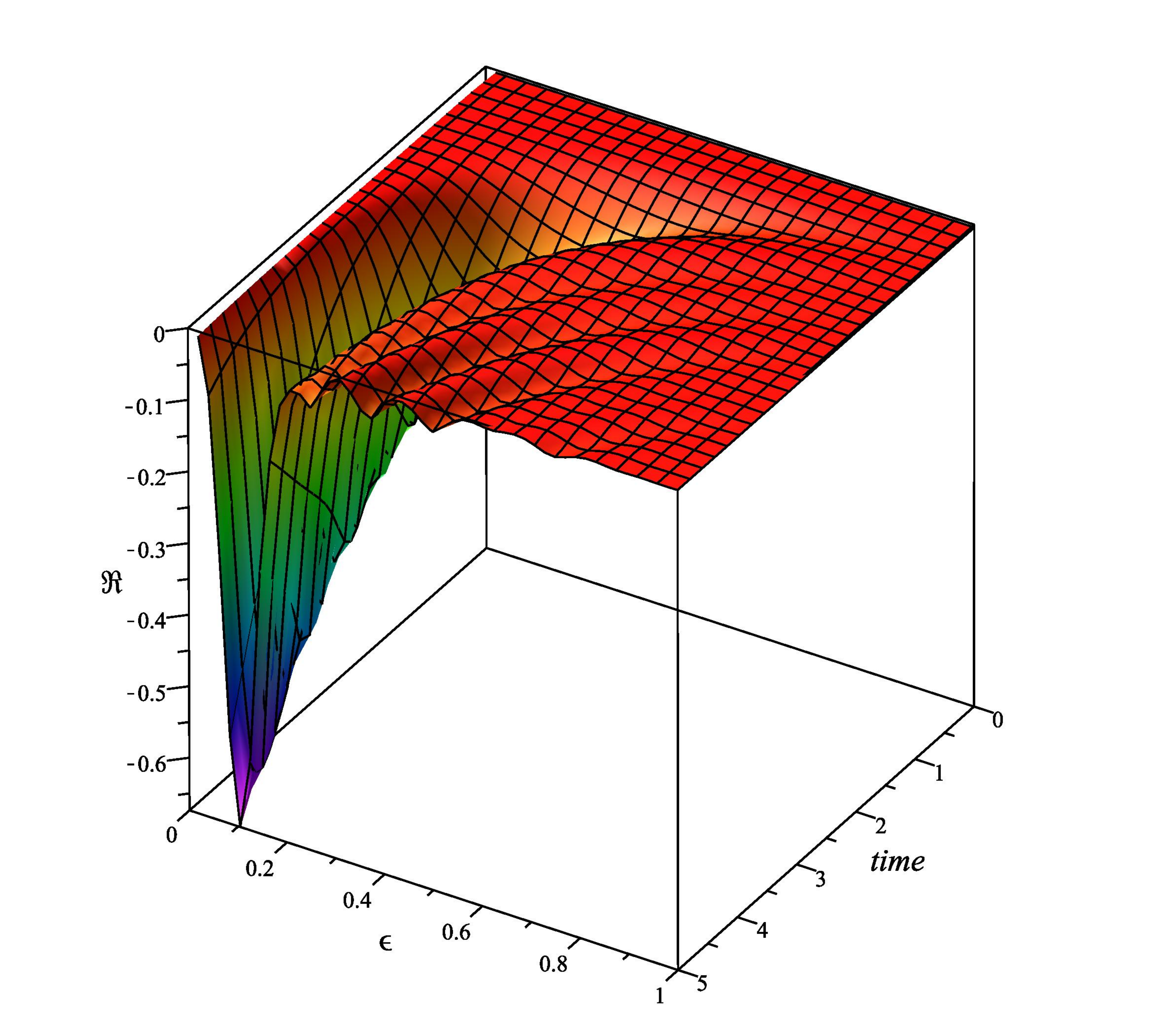}
	\captionof{figure}{\sf (color online) The dynamics of the redistribution factor $ R(t,\epsilon)  $ of entanglement virsus the  time and quench factor $ \epsilon$ for  $ \omega(0)=3, \tilde{\omega}(0)=5, C(0)=1.5 $ and $ \tilde{C}(0)= 0$.}\label{ll}
\end{figure}

In Figure \ref{ll}, we present  the redistribution  factor $ R(t,\epsilon)  $ versus time and quench factor $ \epsilon $ for a specific choice of the coupling parameters and frequencies. The plot  shows that the establishment of  redistribution  phenomenon requires a delay time, which due to Ermakov phases. We notice that by increasing the quench factor $\epsilon$ the delay time decreases dramatically. The redistribution becomes important in the vicinity of the optimal point of entanglement, i.e. $ (t=5,\epsilon=0.1) $ and less important in the regime of time-independent Hamiltonian.  At this level, we switch on the lateral coupling to $ \tilde{C}(0)=4 $  and  by setting the quench factor to unity. 
In that case the covariance matrix takes a simple form because the coefficients $ C_{j}(t)$  of (\ref{eq39}) vanish and consequently the local covariance matrix will be thermal, i.e. diagonal. 

\begin{figure}[htbphtbp]
	\centering 
	\includegraphics[width=5.4cm]{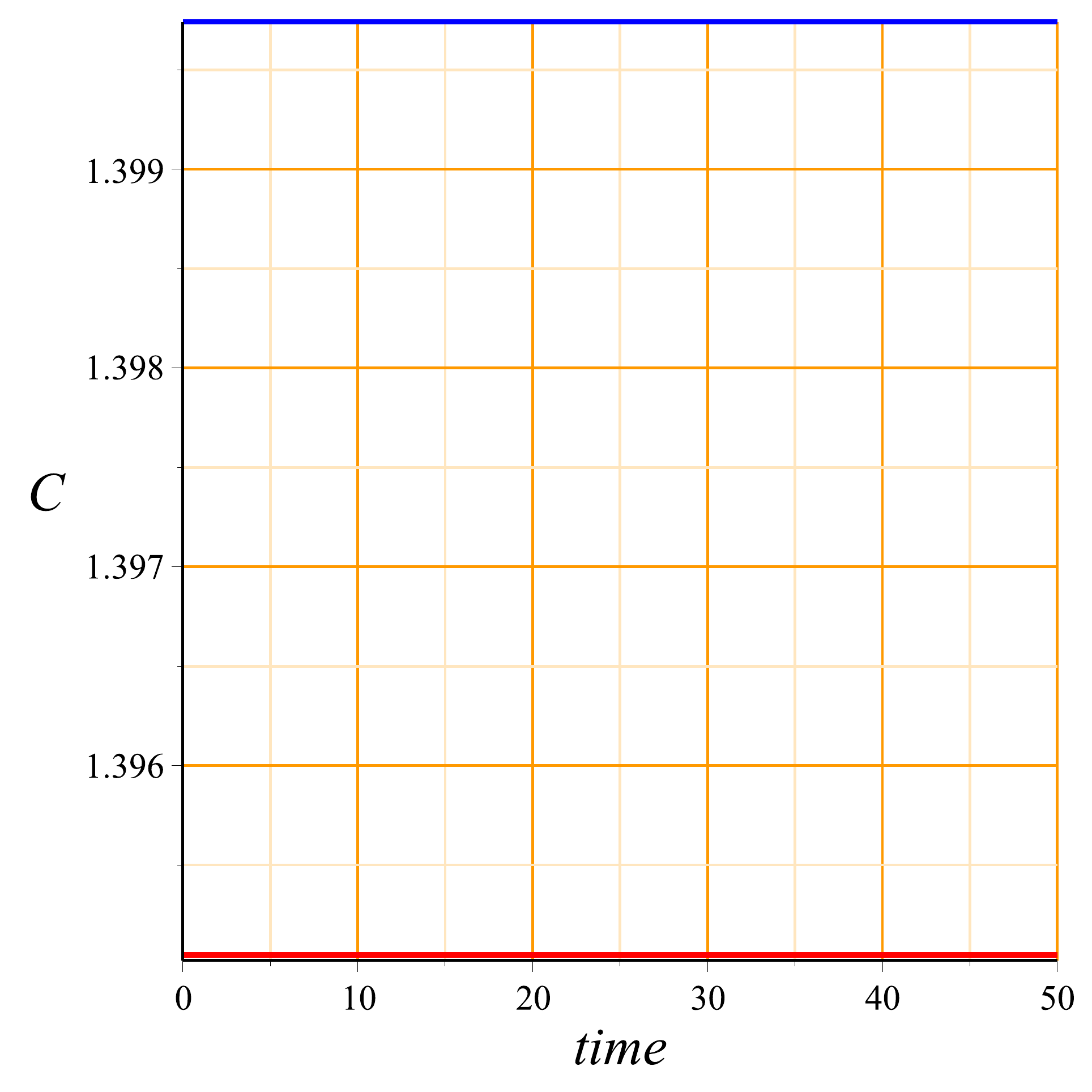} \hspace{2cm}
	\includegraphics[width=5.4cm]{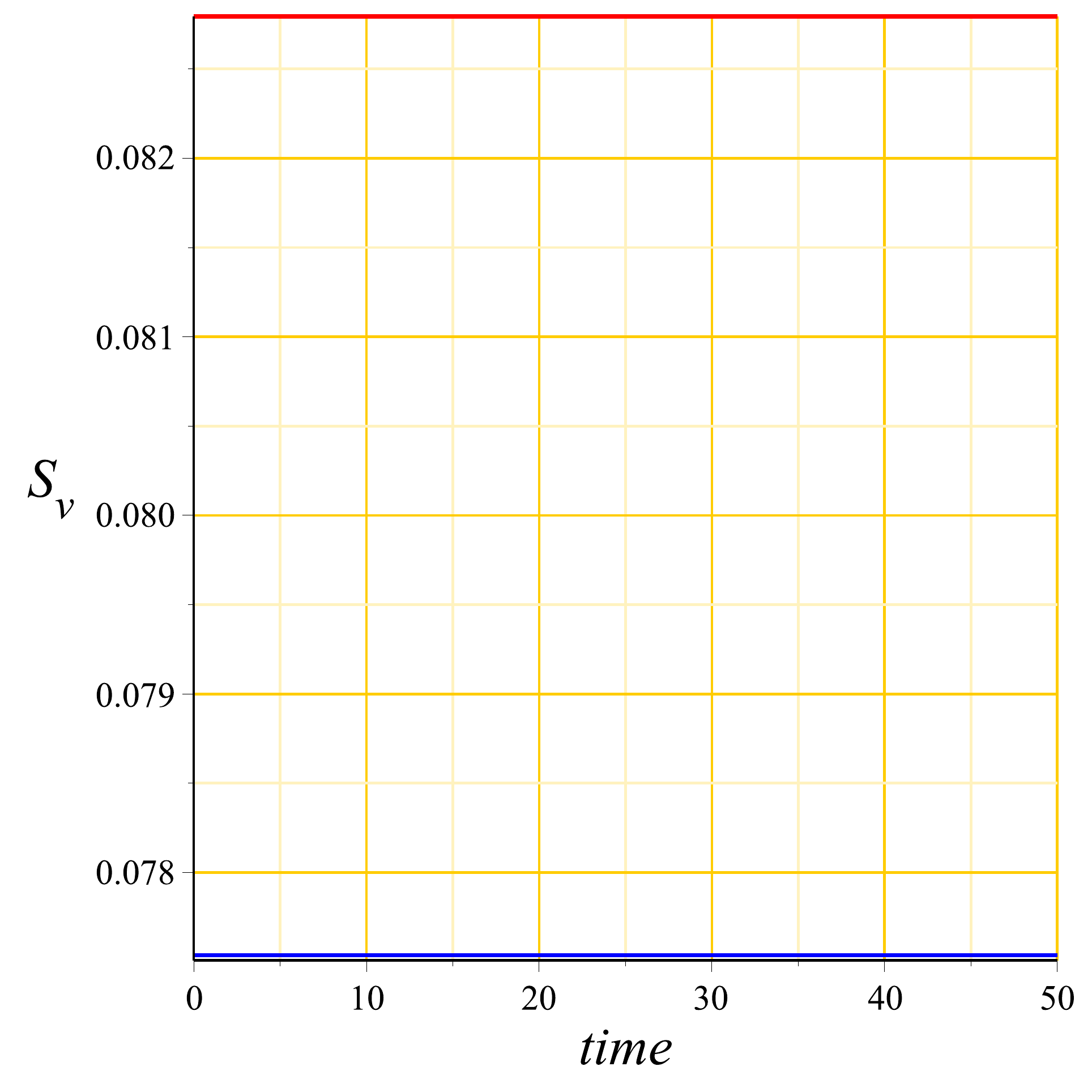}
	\captionof{figure}{\sf (color online)  Entanglement to coherence redistribution  under homodyne detection of the central mode $ B $ (highest frequency mode) for $ \omega(0)=3, \tilde{\omega}(0)=5, C(0)=1.5, \epsilon=1  $ and $ \tilde{C}(0)= 4$. In left panel: coherence of the input state $ \sigma_{A}(in) $ (red solid line) and that of the output state $ \sigma_{A}(out) $ (blue solid line). In right panel: von Neumann entropy $ S_{v} $  of the input state (red solid line) and that of the output state (blue solid line).}\label{homod13}
\end{figure}

In  Figure \ref{homod13} we  observe the consumption of entanglement and the formation of coherence. The opposite process was observed in \cite{R25} by showing that two modes not entangled become entangled after performing a homodyne detection. It is also interesting to note that we have
\begin{equation}
|  C^{A}(out)-C^{A}(in)| \sim |  S_{v}^{A}(out)-S_{v}^{A}(in)|
\end{equation}
which witnesses the redistribution phenomenon.

\section{Conclusion}
We have studied a specific system of interest  namely  three  time-dependent coupled harmonic oscillators following a linear sudden quench (LSQM) dynamics of the coupling parameters and frequencies. The Hamiltonian was diagonalized  by using  a  Euler  time-dependent rotation   together with LSQM, which leads to discard the dynamical effect of  rotation matrix. Later on, we have used the \textit{theorem of eigenvalues-eigenvectors identity} to derive the Euler angles together with rotation matrix inputs.  We have derived the  solution of the time-dependent Shr\"{o}dinger equation of three time-dependent coupled harmonic oscillators.  
Based on the Wigner distribution associated to the vacuum state, the covariant matrix was calculated. 

Subsequently,
we have computed the analytical expressions of the von Neumann entropies and mixedness of each mode in three cases: symmetric, bisymmetric and fully non symmetric. It is shown that the quench factor $\epsilon={\omega_{f}}/{\omega(0)}$ affects strongly the entanglement and coherence amounts. It is noticed that  the  optimal values of quench factor are those near zero, i.e. $\epsilon\longrightarrow 0$. In addition,  we have shown
that the uncertainties and the genuine tripartite entanglement  follow a similar dynamics with respect to the entanglement, which leads to witness their presence in a specific experiment. 
Finally, we have analyzed the  redistribution of entanglement and coherence by using a perfect homodyne detection. It was shown that under some specific choice of the physical parameters the entanglement transforms to coherence and vice verse.



\end{document}